\pdfoutput=1 
\documentclass[a4paper,11pt]{article}

\usepackage{jheppub} 
\usepackage[T1]{fontenc} 

\usepackage{atlasphysics} 
\usepackage{multirow}

\usepackage{preprintcover}  
\PreprintCoverPaperTitle{Measurement of the cross-section for 
    $W$ boson production in association with $b$-jets in $pp$ collisions at
    $\sqrt{s} = 7$ \TeV\ with the ATLAS detector}  
\PreprintIdNumber{CERN-PH-EP-2012-357}  
\PreprintCoverAbstract{This paper reports a measurement of the \Wbjets production cross-section in proton--proton collisions at a centre-of-mass energy of $7 \TeV$ at the LHC. These results are based on data corresponding to an integrated luminosity of $4.6$ \ifb, collected with 
the ATLAS detector. Cross-sections are presented as a function of jet multiplicity and of the transverse momentum of the leading \bjet for both the muon and electron decay modes of the 
$W$ boson. The \Wbjet cross-section, corrected for all known detector effects, is quoted in a limited kinematic range, using jets reconstructed with the anti-$k_t$  clustering algorithm with transverse momentum above $25 \GeV$ and rapidity within $\pm 2.1$.
Combining the muon and electron channels, the fiducial cross-section for \Wbjets  is measured to be 7.1 $\pm$ 0.5 (stat) $\pm$ 1.4 (syst) pb, consistent with next-to-leading order QCD calculations within 1.5 standard deviations.}  
\PreprintJournalName{JHEP}  

\title{\boldmath Measurement of the cross-section for 
    $W$ boson production in association with $b$-jets in $pp$ collisions at
    $\sqrt{s} = 7$ \TeV\ with the ATLAS detector}

\author{ATLAS Collaboration}

\usepackage{hyperref} 
\usepackage[mathlines]{lineno}

\usepackage{xspace}
\def\onejet{1-jet\xspace}
\def\twojet{2-jet\xspace}
\def\threejet{3-jet 1-$b$-tag\xspace}
\def\fourjet{at least 4 jets 1-$b$-tag\xspace}
\def\onejetincl{at least 1 jet 1-$b$-tag\xspace}
\newcommand{\cjets}{{\ensuremath{c}\mbox{-}{\rm jets}}\xspace}
\newcommand{\cjet}{{\ensuremath{c}\mbox{-}{\rm jet}}\xspace}
\newcommand{\bjet}{{\ensuremath{b}\mbox{-}{\rm jet}}\xspace}
\newcommand{\bjets}{{\ensuremath{b}\mbox{-}{\rm jets}}\xspace}
\newcommand{\pTb}{$p_{\textrm T}^{\bjet}$\xspace}

\newcommand{\bhadron}{$b$-hadron\xspace}
\newcommand{\Wbjet}{$W$+\bjets}
\newcommand{\Wbjets}{$W$+\bjets}
\newcommand{\Wcjet}{$W$+\cjets}
\newcommand{\Wljet}{$W$+light-jets\xspace}
\newcommand{\WmT}{$m_{\mathrm{T}}(W)$}

\newcommand{\powheg}  {{\sc Powheg}}

\newcommand{\acermc}  {{\sc AcerMC}}

\newcommand{\mcfm}  {{\sc MCFM}}
\newcommand{\pythia}  {{\sc Pythia}}

\newcommand{\herwig}  {{\sc Herwig}}
\newcommand{\jimmy}  {{\sc Jimmy}}
\newcommand{\alpgen}  {{\sc Alpgen}}

\abstract{This paper reports a measurement of the \Wbjets production cross-section in
proton--proton collisions at a centre-of-mass energy of $7 \TeV$ at the LHC. These results are based on data corresponding to an integrated luminosity of $4.6$ \ifb, collected with 
the ATLAS detector. 
Cross-sections are presented as a function of jet multiplicity and of the transverse
momentum of the leading \bjet for both the muon and electron decay modes of the 
$W$ boson. 
The \Wbjet cross-section, corrected for all known detector effects, is quoted in a limited kinematic range, using jets reconstructed with the anti-$k_t$  clustering algorithm with transverse momentum above $25 \GeV$ and rapidity within $\pm 2.1$.
Combining the muon and electron channels, the fiducial cross-section for \Wbjets  is measured to be
7.1 $\pm$ 0.5 (stat) $\pm$ 1.4 (syst) pb, consistent with next-to-leading order QCD calculations within 1.5 standard deviations.
}

\begin{document} 
\maketitle
\flushbottom

\section{Introduction}

This paper reports a measurement of the cross-section for $W$ boson production in association with $b$-quark jets in proton--proton ($pp$) collisions at $\sqrt{s}$ = 7 \TeV. The measurement, performed differentially in the \bjet transverse momentum (\pT) for the 1-jet and 2-jet final states, provides an important test of perturbative quantum chromodynamics (QCD) in the presence of heavy quarks. 

Next-to-leading-order (NLO) perturbative QCD calculations of the \Wbjets process  have recently become available in Monte Carlo (MC) simulations, both at the parton level~\cite{MCFM} and enhanced with parton shower models~\cite{Powheg, MC_at_NLO}. Several processes contribute to \Wbjets production at NLO. In the four-flavour number scheme (4FNS), where only $u$, $d$, $c$, $s$ are considered as initial-state quarks, these are 
$q\qbar \to W b\bbar(g)$ and 
$gq \to W b\bbar q$.
When considering the presence of $b$-quarks in the initial state (5FNS), the single $b$-quark processes
$bq \to W bq(g)$ and 
$bg \to W bq\qbar$
also play a significant role at LHC energies~\cite{ref:4FNS5FNSref, ref:4FNS5FNS}.
In addition, double-parton interactions (DPI), in which a $W$ boson and \bjets are produced from different parton--parton interactions within the same $pp$ collision, are also expected to contribute to the total observed \Wbjets cross-section~\cite{MPI_WBB}. 

The production of \Wbjets events via top-quark decay (from single or pair-produced top-quarks) is not included in the signal definition for the primary measurement. An additional set of measurements is  performed including the contribution from single top-quark production. These are of particular relevance for the differential cross-section, since single top-quark  and \Wbjet events are difficult to separate and their relative contribution depends strongly on the \bjet transverse momentum.  

The \Wbjets process is an important background to the Higgs boson associated-production process $WH$ with $H\to\bbbar$ decays~\cite{WH}. The associated-production measurements are a substantial ingredient in determining the coupling of the Higgs boson to fermions, through the decay $H\to\bbbar$, and searches in this channel have been included in studies of the Higgs-like boson~\cite{HiggsComb}. 
The \Wbjets process is also an irreducible background in some searches for physics beyond the Standard Model~\cite{NewPhys1}, and in measurements of single top-quark properties~\cite{SingleTop}, due to the dominating branching fraction of the $t \to Wb$ decay. 

Measurements of the \Wbjets fiducial cross-section in proton--antiproton collisions
at $\sqrt{s} = 1.96 \TeV$ have been reported by the CDF Collaboration~\cite{Aaltonen:2009qi} and more recently also by the D0 Collaboration ~\cite{D0Wb}.
The ATLAS Collaboration reported a previous measurement based on 36~\ipb\  of data collected in $pp$ collisions at $\sqrt{s} = 7 \TeV$~\cite{bib:ATLASWb}. 
The CDF measurement of $2.74 \pm 0.27~(\rm{stat}) \pm 0.42~(\rm{syst})$~pb, and the ATLAS measurement of
$10.2 \pm 1.9~(\rm{stat}) \pm 2.6~(\rm{syst})$~pb are both found to be larger than the corresponding theoretical cross-sections calculated at NLO ($1.22 \pm 0.14$~pb and $4.8\pm1.3$~pb) by 2.8 and 1.5 standard deviations, respectively.
The D0 measurement of $1.05 \pm 0.12$~pb is found to be lower than the theoretical prediction of $1.34^{+0.41}_{-0.34}$~pb~\cite{MCFM}, but in agreement within theoretical uncertainties.

In this paper, the \Wbjets cross-section   is measured using the ATLAS detector\footnote{The ATLAS experiment uses a right-handed coordinate system
  with its origin at the nominal interaction point (IP) in the centre
  of the detector and the $z$-axis along the beam pipe. The $x$-axis
  points from the IP to the centre of the LHC ring, and the $y$-axis
  points upward. Cylindrical coordinates $(r,\phi)$ are used in the
  transverse plane, $\phi$ being the azimuthal angle around the beam
  pipe. The pseudorapidity is defined in terms of the polar angle
  $\theta$ as $\eta=-\ln\tan(\theta/2)$ and the rapidity is defined as
  $y = \ln[(E+p_z)/(E-p_z)]/2$.  The distance $\Delta R$ in
  $\eta$-$\phi$ space is defined as $\Delta R\equiv \sqrt{(\Delta\eta)^2
    + (\Delta\phi)^2}$.} in a restricted fiducial region defined at the particle level and given in table~\ref{tab:fiducialps}. For the first time, and in the same region, the \Wbjets differential cross-section is also measured as a function of the \bjet $\pT$.
\begin{table}[ht]
\centering
\caption{Definition of the phase space for the 
  fiducial region. The $W$ transverse mass is defined as
$ m_\mathrm{T}(W) = \sqrt{2p^{\ell}_\mathrm{T} p^{\nu}_\mathrm{T} (1 -
  \cos(\phi^{\ell} - \phi^{\nu})) }$. 
}
\begin{tabular}{l|c}
\hline\hline
Requirement & 
Cut\\ 
\hline \hline
Lepton transverse momentum & $\pT^\ell > 25$ \GeV  \\
Lepton pseudorapidity & $|\eta^\ell|<2.5$  \\
Neutrino transverse momentum &   $\pT^\nu >25$ \GeV \\
$W$ transverse mass & $m_\mathrm{T}(W)>60$ \GeV \\
\hline
Jet transverse momentum& $\pT^j>25$ \GeV \\
Jet rapidity &  $|y^j|<2.1$   \\
Jet multiplicity & $n \le 2$ \\
$b$-jet multiplicity & $n_b = 1$ or $n_b = 2$\\
\hline
Jet-lepton separation & $\Delta R(\ell\mathrm{,jet}) > 0.5$  \\
\hline
\hline
\end{tabular} 
\label{tab:fiducialps}
\end{table}
To enter the fiducial region, events at the generator level are required to contain an electron or muon and a neutrino originating from a $W$ boson decay, and one or two hadron-level jets%
\footnote{Hadron-level jets are built from stable particles, i.e. those with a proper lifetime longer than 10 ps. This definition includes muons and neutrinos from decaying hadrons.
}. 
At least one of the jets is required to be a \bjet, defined by the presence of a weakly decaying \bhadron\ with $\pT > 5$ GeV and within $\Delta R = 0.3$ of the jet axis. 

At the reconstruction level, events are required to be consistent with the decay of a $W$ boson to the $\ell \nu$ ($\ell = \mu, e$) final state, and to contain either one or two jets.  Events are selected if exactly one jet, which can be a \bjet, but also a mis-tagged \cjet or a light-jet, passes the $b$-tagging requirements. Events with two or more $b$-tagged jets are rejected, as are events with three or more jets, to reduce the top-quark background.
Given the fiducial region definition and the reconstruction-level selection,  the measurement is performed using reconstructed events containing a single $b$-tagged jet, and unfolded to the fiducial region with one or more \bjets. 
The measurement is performed separately in the $W\to \mu\nu$ and $W\to e\nu$ decay channels and in the exclusive 1-jet and 2-jet final states.

\section{The ATLAS detector} 
The ATLAS detector~\cite{bib:JINST} is a multi-purpose particle physics detector 
operating at one of the interaction points of the LHC.
It consists of an inner detector tracking system (ID) within a $2$~T axial magnetic field
provided by a superconducting solenoid, surrounded by electromagnetic and 
hadronic calorimeters, and by a muon spectrometer (MS) embedded in the magnetic field provided by three 
air-core superconducting toroidal magnets.  

The ID consists of pixel and silicon microstrip detectors surrounded by a transition radiation
tracker, and it provides measurements of charged-particle tracks within $|\eta| < 2.5$.  
The calorimeters provide three-dimensional reconstruction of particle showers in the region
of $|\eta| < 4.9$, with a finely segmented inner layer used for electron identification in $|\eta| < 2.5$.
They are based on liquid-argon (LAr) sampling technologies, except for the barrel 
region ($|\eta| < 1.7$) of the hadronic calorimeter where scintillator tiles are used
as the active media.
The MS consists of three layers of high-precision tracking chambers (monitored drift-tubes
and cathode strips) in the region $|\eta| < 2.7$, and resistive-plate or thin-gap chambers  providing trigger signals in the region $|\eta| < 2.4$.

\section{Simulated event samples}
Monte Carlo simulated samples are
used to model the reconstructed $W$+$b$-jet signal and most of its background contributions, as
well as to extract a fiducial cross-section from the measured $W$+$b$-jet yield. 

The processes of $W$ boson production in association with $b$-jets,
$c$-jets and light-jets are simulated separately using the
\textsc{Alpgen}~2.13~\cite{ALPGEN} generator, interfaced to
\textsc{Herwig}~6.510~\cite{Herwig} for parton showers and hadronization, and
\textsc{Jimmy}~4.31~\cite{Jimmy} for the underlying-event simulation. 
Exclusive samples with zero to four additional partons and an inclusive sample with five or more additional partons are used.
The MLM~\cite{MLM} matching scheme, as implemented in \textsc{Alpgen}, is
used to remove overlaps between samples with the same parton multiplicity
originating from the matrix element (ME) and the parton shower (PS).
  In addition,
overlap between samples with heavy-flavour quarks originating from the ME and 
 from the PS is removed.
Large samples of dijet events simulated using  \pythia{}~6.423~\cite{Sjostrand:2006za}
are used to model the light and heavy-flavour jet properties relevant to the \bjet identification in the $W$+jets sample.

The $Z$+jets background is simulated
with \alpgen\ interfaced to \textsc{Herwig} and \textsc{Jimmy}, using the same
configuration as for $W$+jets. The diboson ($WW$, $WZ$, $ZZ$)
background is simulated with \textsc{Herwig}. 
The $t$-channel, $s$-channel and $Wt$-channel single-top processes 
are simulated with AcerMC~3.7~\cite{bib:ACERMC} interfaced to {\pythia}.
The \ttbar\ background is simulated with \powheg~\cite{Powheg} interfaced to \textsc{Pythia}.

The total cross-sections of the  $W$+jets
and $Z$+jets samples are normalized to the inclusive NNLO
predictions~\cite{FEWZ}, while other backgrounds are normalized to NLO predictions~\cite{VVnormalization, SingleTopNormalization, TopNormalization}.  
The \ttbar\ contribution in the 1-jet and 2-jet analysis regions, and the 
single-top contribution in the 2-jet analysis region, are estimated from data. 

For all the processes modelled, multiple interactions
per bunch crossing (pile-up) are accounted for by overlaying minimum-bias events simulated
with \pythia{}  onto the generated hard process. The detector simulation~\cite{AtlasSimulation} 
is based on the \textsc{Geant}4 program~\cite{Geant4}.

\section{Data sample and event selection}\label{sec:selection}
The analysis considers data recorded in the year 2011 during periods with stable $pp$ collisions at $\sqrt{s} = 7 \TeV$,
and where all relevant parts of the detector were operating normally. The resulting data set corresponds to 4.6 \ifb\ 
of integrated luminosity, with an uncertainty of 3.9\%~\cite{bib:Lum1, bib:Lum2}.
Events were collected using single-muon or single-electron triggers. The \pT\ threshold of the muon trigger was 18 \GeV, while the transverse energy ($\ET$) threshold used for the electron trigger was initially 20 \GeV\ and was later raised to 22 \GeV\ to cope with the increasing LHC instantaneous luminosity.

Candidate \Wbjets events are required to have exactly one high-\pt\ electron or muon, as well
as  missing transverse momentum ({\MET}) consistent with a neutrino from a $W$ boson,
and one or two reconstructed jets, exactly one of which must be $b$-tagged.
All events must have at least one reconstructed vertex formed by the intersection
of at least three tracks with $\pt> 400 \MeV$. In events with multiple vertices, the
vertex with the largest sum of squared \pt\ of the associated tracks is taken to be 
the primary hard-scatter vertex (PV).

Requiring events to have exactly one $b$-tagged jet significantly reduces the top-quark background contribution in the \twojet analysis region.
\Wbjets events with a second \bjet satisfying the fiducial selection represent 10\% of the 2-jet fiducial region. Most of these events have a single $b$-tagged jet, and they are included in the \twojet region at the reconstruction level.  

Electron candidates are formed by matching clusters found in the electromagnetic calorimeter to tracks reconstructed in the ID
in the region of  $|\eta| < 2.47$ and are required to have $\ET > 25\GeV$.  To ensure good containment of  electromagnetic showers in the calorimeter,  the 
 transition region $1.37< |\eta| <1.52$ between the barrel and the endcaps is excluded. The lateral and transverse shapes of the clusters must be consistent with those of an electromagnetic shower~\cite{TagProbeEle}.
Muon candidates, reconstructed by combining tracks reconstructed in the ID and the MS,
are selected in the region $|\eta| < 2.4$, and are required to have $\pT > 25\GeV$. 
Both the electron $\ET$ and the muon \pt\ requirements are chosen to be on the efficiency plateau for the respective triggers. 
The selection efficiency of electrons and muons in simulated events, as well as their energy and 
momentum scale and resolution, are adjusted to reproduce those observed in $Z\to \ell\ell$ events in data~\cite{MomentumScaleMu,TagProbeMu,TagProbeEle}.

In order to reduce the large background from multijet production, 
lepton candidates are required to be isolated from neighbouring tracks
within $\Delta R = 0.4$ of their direction, as well as from other calorimeter energy 
depositions, corrected for pile-up contributions, within $\Delta R = 0.2$. 
In the muon case, the sum of transverse momenta of neighbouring tracks must be less
than $2 \GeV$, while the sum of the calorimeter transverse energies must be less 
than $1 \GeV$.  
In the electron case, these requirements
range between 1.35~\GeV\ and 3.15~\GeV\ depending on $\pT$ and $\eta$ in order to yield a constant efficiency across momentum ranges and detector
regions.
Additionally, leptons are required to be consistent with originating from the PV. 
Their longitudinal impact parameter ($|z_0|$) with respect to the PV  must be smaller 
than 10~mm, and the ratio of the transverse impact parameter $d_{0}$ to its uncertainty ($d_{0}$ significance)
must be smaller than 3 for muons, and 10 for electrons.

Jets are reconstructed from calorimeter energy topological clusters (topoclusters)~\cite{TopoClusters} using the anti-$k_t$ 
algorithm~\cite{ref:antikt} with a radius parameter $R=0.4$. 
They are required to have a transverse momentum greater than 25~\GeV, and a rapidity 
 $|y| < 2.1$ in order for the entire jet to be reconstructed within the tracking region.
 Jets originating in pile-up interactions are suppressed 
by requiring that at least 75\% of the total transverse momentum of tracks associated with each jet
 point to the PV.  
Jets within a distance $\Delta R = 0.5$ of the lepton candidate are removed, and jets 
arising from detector noise or cosmic rays are also rejected~\cite{JetCleaning}. 

The jet energy is calibrated to account for the different response 
of the calorimeters to electrons and hadrons, for energy losses 
in un-instrumented regions, and for the energy offset introduced by pile-up,
by applying jet calibration factors dependent on $\pT$, $\eta$~\cite{ref:JES, jeseta, jes}, and pile-up conditions~\cite{JESpileup}. 
A residual calibration derived from in-situ techniques is applied to the data to reduce
differences between data and Monte Carlo simulation~\cite{JESInSituZ, JESInSituGamma}.

Jets originating from $b$-quarks are identified using the combination of two $b$-tagging algorithms.
The first one exploits the topology of weak $b$- and $c$-hadron decays inside the jet to reconstruct their 
decay vertices along a common line originating from the PV. 
The second one uses the impact parameter significance of each track within the jet to determine the likelihood  that the jet originates from a $b$-quark.
The properties measured by these two taggers are combined using an artificial neural network to determine a single discriminant variable (CombNN)~\cite{bib:AdvancedTaggers}.

The CombNN variable is used both to select a sample enriched with \bjets, as well as to discriminate between $b$-jets, $c$-jets and light-jets within the enriched sample.
The  working point used for the selection (CombNN$~ > 2.2$) corresponds to a $b$-tagging efficiency 
of about 40\% at low \pt, increasing to a plateau of 57\% for $b$-jets of $\pt$ above 60 \GeV,
with rejection rates of about 10 for $c$-jets and 1000 for light-jets.
In order to reproduce the $b$-jet, $c$-jet and light-jet tagging efficiencies measured in data~\cite{ref:BtaggingSF, ref:BtaggingSFC, ref:BtaggingSFL}, event weights in Monte Carlo simulation are scaled as a function of the number of tagged and untagged jets of each flavour and the corresponding \pT\ and $\eta$.
The CombNN distribution in the $b$-tagged sample is then used to separate statistically the remaining $c$-jet and light-jet contributions from the \Wbjets signal.

The measurement of \met\ in each event is based on an algorithm~\cite{ref:METRefFinal} which performs the vector sum of transverse energies of  high-$\pT$ objects such as electrons, muons and jets, and of  individually calibrated~\cite{Barillari:2009zza} topological energy clusters~\cite{TopoClusters} not associated with any physics objects. 
To be consistent with a $W$ boson decay, and to reduce the multijet background, the \met\ is required to be larger than $25\GeV$, and the   $W$ boson transverse mass \WmT{}~$= \sqrt{2p^{\ell}_\mathrm{T} p^{\nu}_\mathrm{T} (1 - \cos(\phi^{\ell} - \phi^{\nu})) }$
is required to be larger than $60\GeV$.

\section{Signal and background estimation}
Several processes contribute to the overall background for the \Wbjet signal,  accounting for more than 85\% of the selected sample. Some of the backgrounds, such as single-top, \ttbar\ and multijet, are characterized by rather different kinematics from \Wbjet, but they have real \bjets in their final state and show similar $b$-tagging response. Others,  \Wcjet and \Wljet, have kinematic properties similar to the signal, but they can be statistically separated by studying the characteristics of $b$-tagged jets. The remaining backgrounds, diboson ($WW$ and $WZ$) and $Z$+jets, contribute less than 5\% of the selected sample.

In most cases, backgrounds are estimated directly from data in order to reduce the theoretical uncertainties on their normalization.
A sequence of binned maximum likelihood (ML) fits is performed, in which a distribution measured in data is described by a linear combination of templates representing each contributing process. In each fit, the normalization of the process of interest is allowed to float freely, while some processes are constrained by Gaussian terms in the 
likelihood\footnote{Using log-normal terms to ensure positive normalization yields consistent results with those obtained using Gaussian terms.}.  For a constrained process, the mean of the corresponding Gaussian constraint is fixed to the expected number of events, while the width is fixed to the associated uncertainty. This uncertainty is derived either from the results of a previous ML fit or from theoretical uncertainties, depending on the process considered. 
Pseudo-experiments are used to validate the behaviour and properties of each fit. 

The single-top, \ttbar\ and multijet contributions are estimated either in background-enriched control regions or using kinematic distributions directly in the signal regions. The \Wbjet,  \Wcjet and \Wljet contributions are then statistically separated, and the number of \Wbjet events is extracted, by fitting the CombNN weight distribution of $b$-tagged jets observed in data in each analysis region. Example templates for the muon \onejet sample are shown in figure~\ref{fig:OverCombNN1}.

\begin{figure*}[h]
\begin{center}
\includegraphics[width=0.6\textwidth]{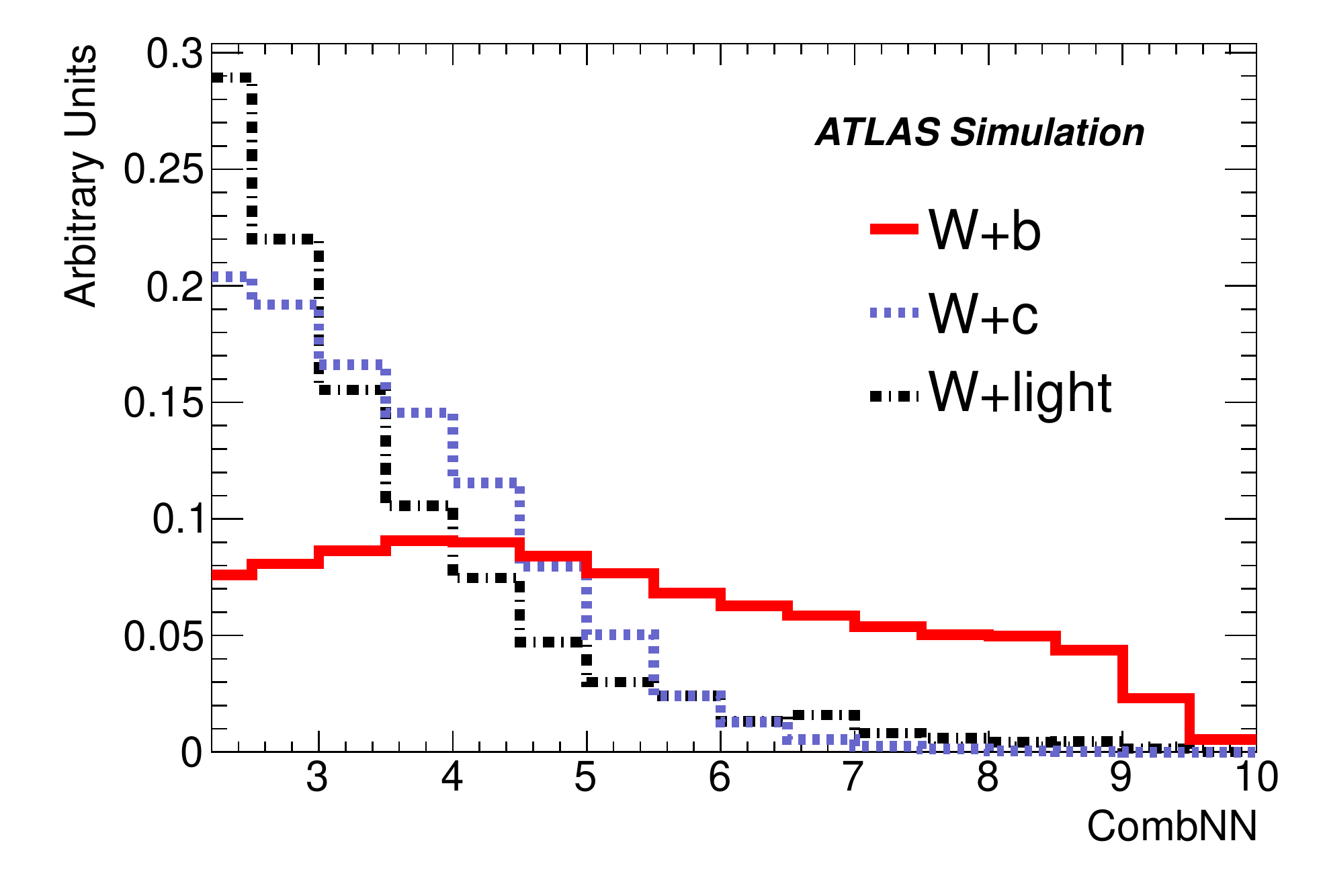}
\caption{ Overlay of the \Wbjet, \Wcjet and \Wljet CombNN distributions in MC simulation for $b$-tagged (CombNN $ > 2.2$) jets in the muon \onejet analysis region.}
\label{fig:OverCombNN1}
\end{center}
\end{figure*}

\subsection{Multijet background}\label{sec:Multijet_Background}

Multijet events from QCD production processes in which one of the jets is either identified as a lepton, or contains a real lepton originating from a heavy-quark decay, can occasionally enter the selected sample. Specific criteria are used to reduce this contamination, such as the lepton identification, isolation and impact parameter, and the \met{} and \WmT{} requirements mentioned in section~\ref{sec:selection}. To estimate the remaining multijet contribution, complementary data samples highly enriched in multijet events are created by requiring that some of these criteria are not fulfilled. The normalization of these samples is then obtained by fitting the \met{} or \WmT{} distributions in data.

Specifically, the multijet background shape for each distribution is obtained in the muon channel by inverting the tracking isolation requirement, and in the electron channel by inverting part of the identification selection and waiving the calorimeter isolation requirement. 
The selection used to form the multijet template from data was studied using dijet Monte Carlo simulations to minimize kinematic biases with respect to the standard signal selection, while maintaining the large number of events required to obtain smooth templates. 

The normalization of the multijet template is then assessed, in each analysis region, by performing a fit to the \met{} distribution in data after relaxing the \WmT{} requirement from  60~\GeV{} to 40~\GeV{}, and removing the \met{} $> 25 \GeV$ requirement. 
The templates used in this fit for the $W/Z$+jets, \ttbar, single-top and diboson processes are based on Monte Carlo simulation. The multijet and the $W$+jets template normalizations are free parameters of the fit to the \met{} distribution, while those of the other components are fixed to their expected cross-sections.
The $\met$ distributions, normalized to the results of the fit, are presented in figure~\ref{fig:QCD_METfit1} for the 1- and 2-jet regions in the muon and electron channels.  

\begin{figure*}[ht]
\begin{center}
\includegraphics[width=0.49\textwidth]{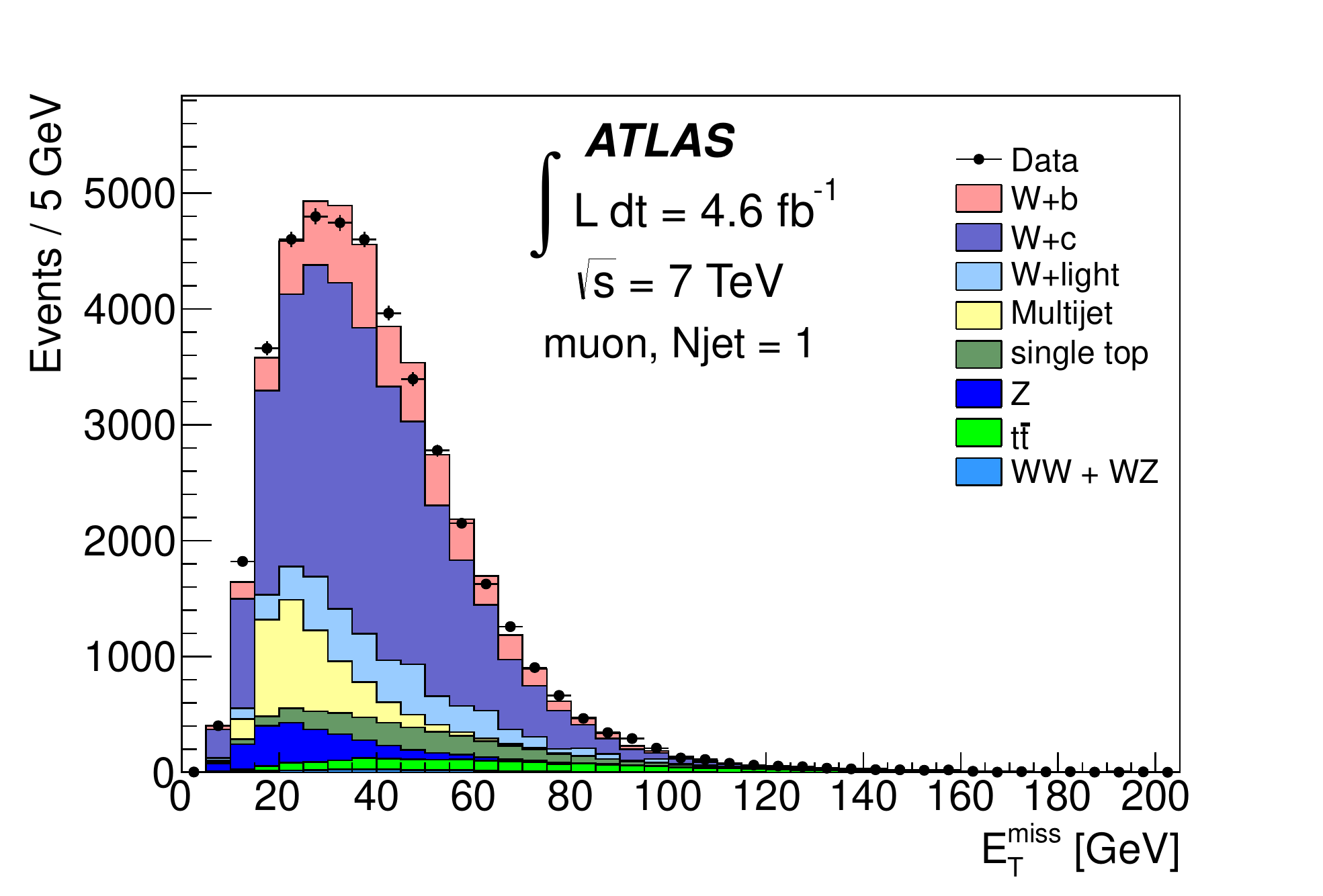}
\includegraphics[width=0.49\textwidth]{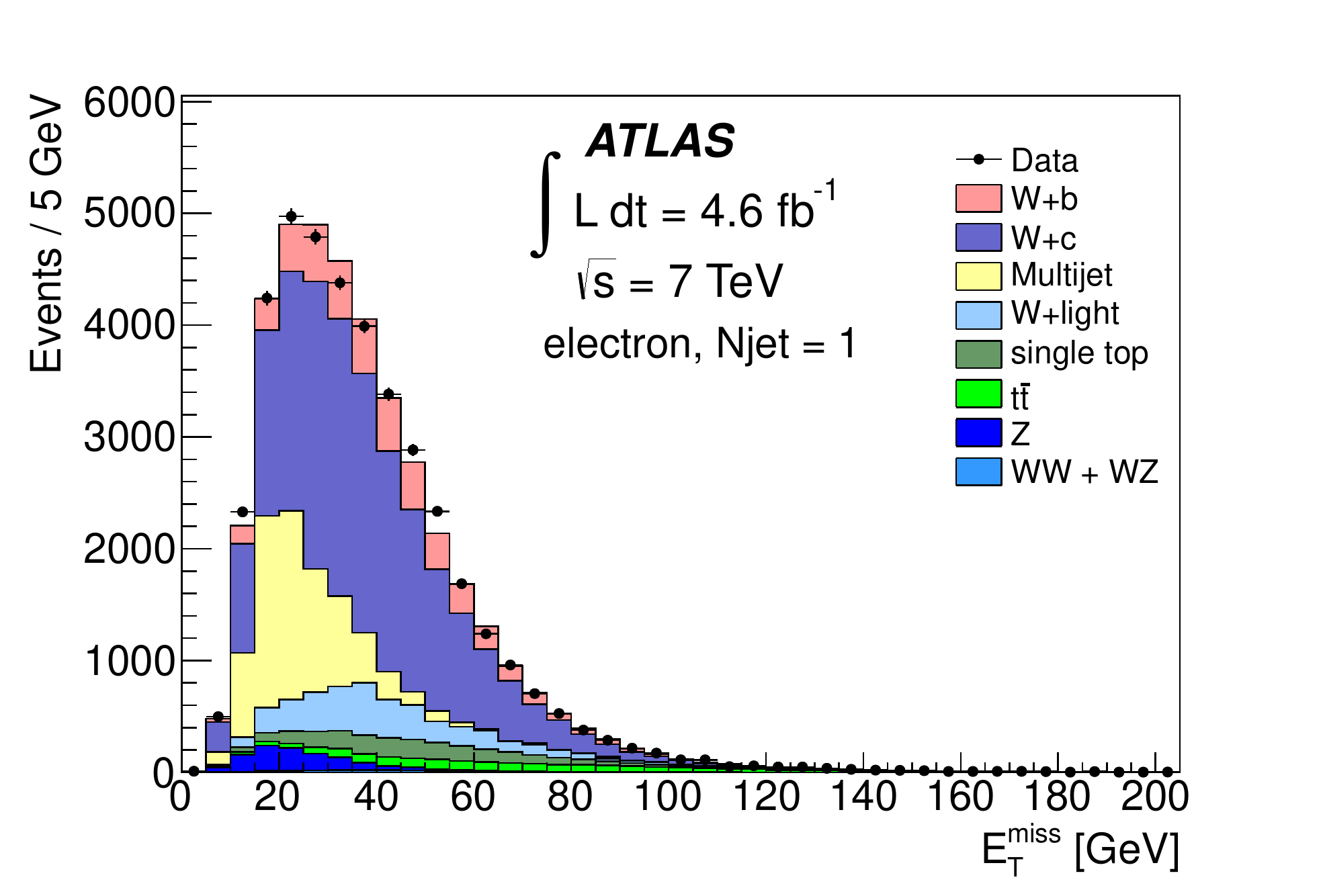}
\includegraphics[width=0.49\textwidth]{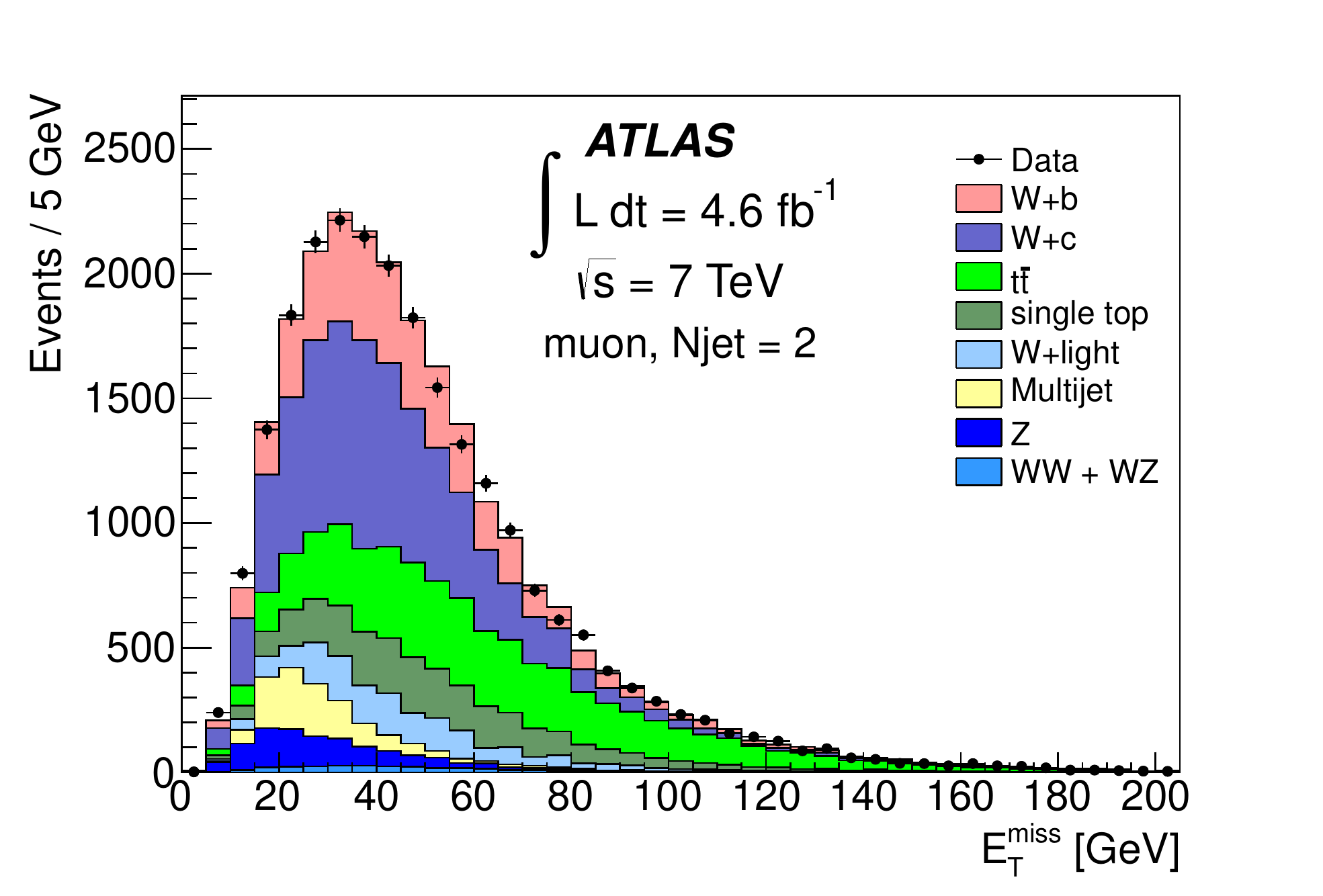}
\includegraphics[width=0.49\textwidth]{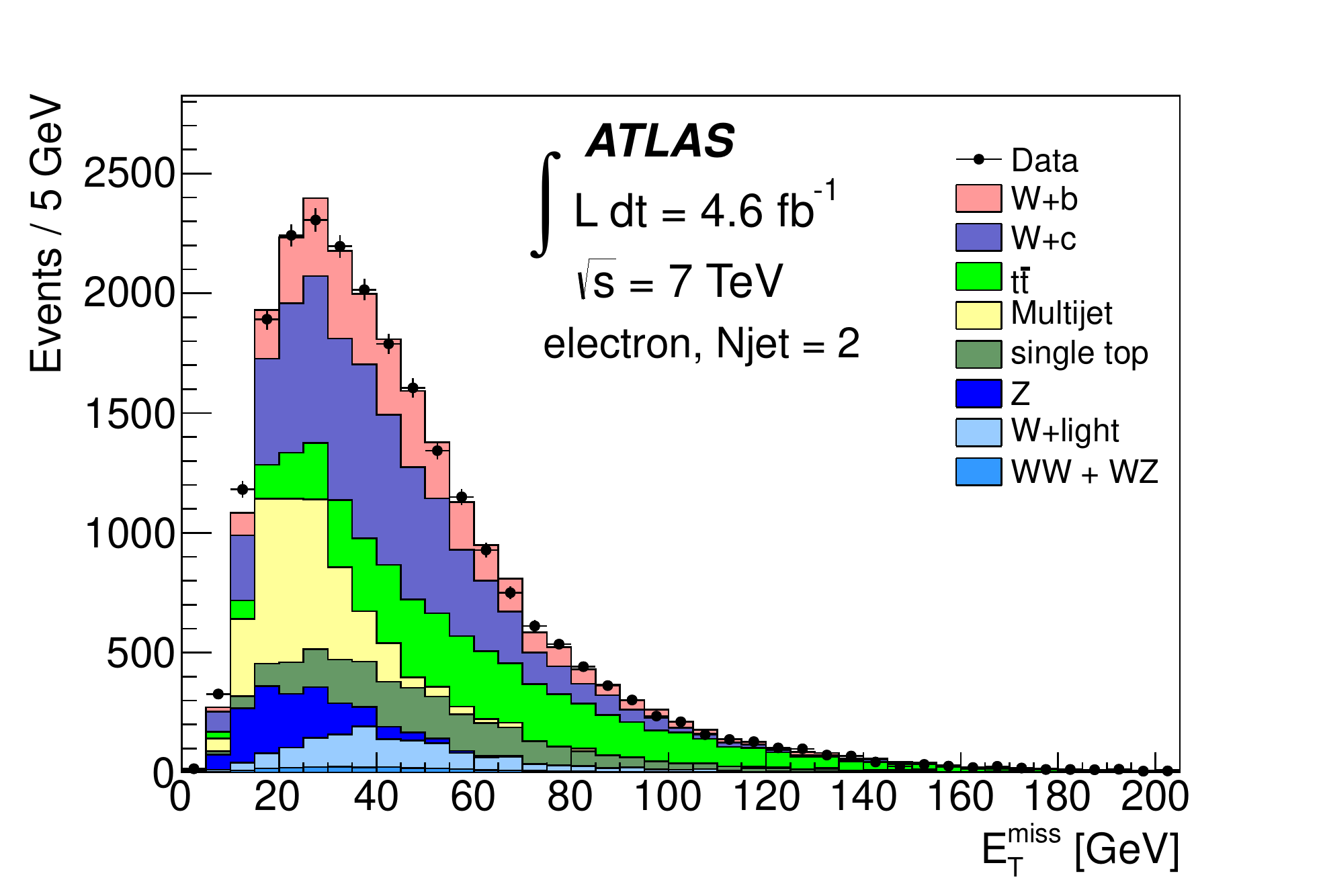}
\caption{  \met{} distributions in data and MC simulation in the \onejet (top) and \twojet (bottom) analysis regions, in the muon (left) and electron (right) channels. MC samples are normalized to the results of the multijet background fit. To enhance the multijet contribution in the fitted region, the  \WmT{} selection is loosened from 60~\GeV{} to 40~\GeV{}.}
\label{fig:QCD_METfit1}
\end{center}
\end{figure*}

An uncertainty of 50\%, applied as a Gaussian constraint in subsequent ML fits, is assigned to the multijet normalization by comparing the \MET{} fit results in each analysis region with the results obtained by fitting the alternative distributions of \WmT{}  and  lepton \pt{}.  In the 4-jet region used to estimate the \ttbar{} background, this uncertainty is estimated to be 100\%.
In the differential measurement, the multijet background normalization is extrapolated from the inclusive estimates, and the same 50\% uncertainty is applied as an independent Gaussian constraint in each \pTb bin.

\subsection{$t\bar{t}$ background}\label{sec:tt_Background}

The \ttbar\ background is estimated in data by selecting events with at least four jets and exactly one $b$-tag. A binned ML fit to the CombNN distribution 
is performed in this control region to extract the \ttbar\ yield. The \ttbar\ Monte Carlo simulation is then used to extrapolate the measured yield into the 1- and 2-jet analysis regions.

In this fit, the sum of the \Wbjet, \Wcjet and \Wljet MC templates is normalized to the NNLO $W$ inclusive cross-section. Their relative contributions are taken from the \alpgen\ Monte Carlo prediction and a Gaussian normalization uncertainty constraint of 100\% is applied to each. Similarly, the single-top template is assigned a 50\% constraint that reflects the maximum uncertainty on the single-top normalization discussed in section~\ref{sec:SingleTop_Background}. The multijet background is estimated using the technique described in the previous section and assigned a normalization uncertainty constraint of 100\%, based on the fits to the alternative distributions. The $Z$+jets  contribution is assigned a 10\% normalization uncertainty constraint based on theoretical calculations and previous measurements \cite{ztheo, zjetatlas}. Finally, the diboson contribution is assigned a 10\% normalization uncertainty constraint, which is twice the uncertainty of 
the corresponding NLO predictions~\cite{wwz1, wwz2}.

The correction factors to the \ttbar\ Monte Carlo normalization estimated by the fit in the ``\fourjet'' region are $1.09\pm0.06$ for the muon channel and $1.08\pm0.07$ for the electron channel.  These factors are in good agreement with those resulting from alternative fits to the ``\threejet'' region, and with those resulting from a fit of the number-of-jets distribution in the ``\onejetincl'' and the ``\fourjet'' regions. The fit projections corresponding to the CombNN distribution in the ``\fourjet'' region and to the number-of-jets distribution in the ``\onejetincl'' are shown in figure~\ref{fig:ttcontrolfit}. As a result, a 10\% \ttbar\ normalization uncertainty is applied as a Gaussian constraint in  subsequent fits. 
 
\begin{figure*}[!htb]
\begin{center}
\includegraphics[width=0.49\textwidth]{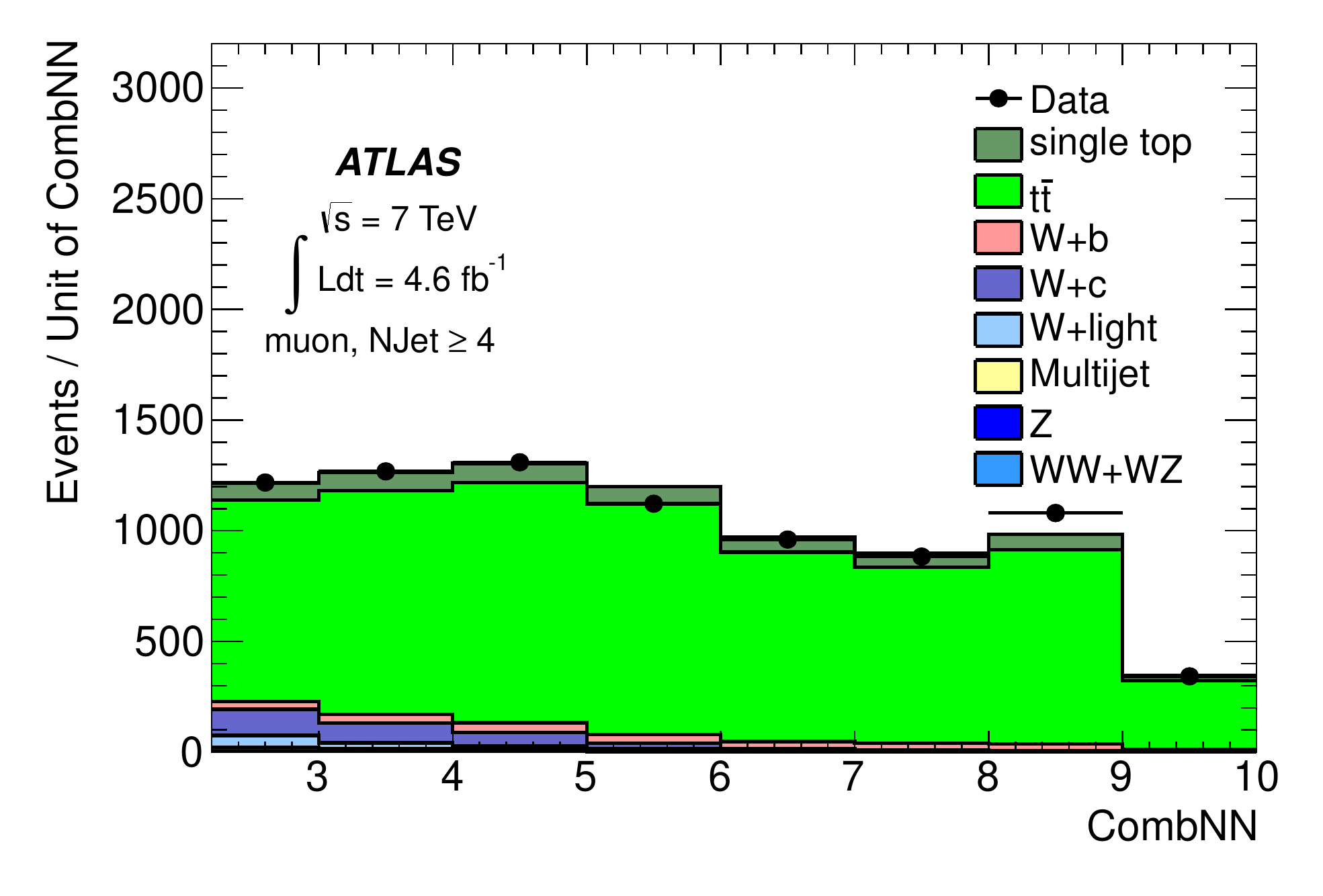}
\includegraphics[width=0.49\textwidth]{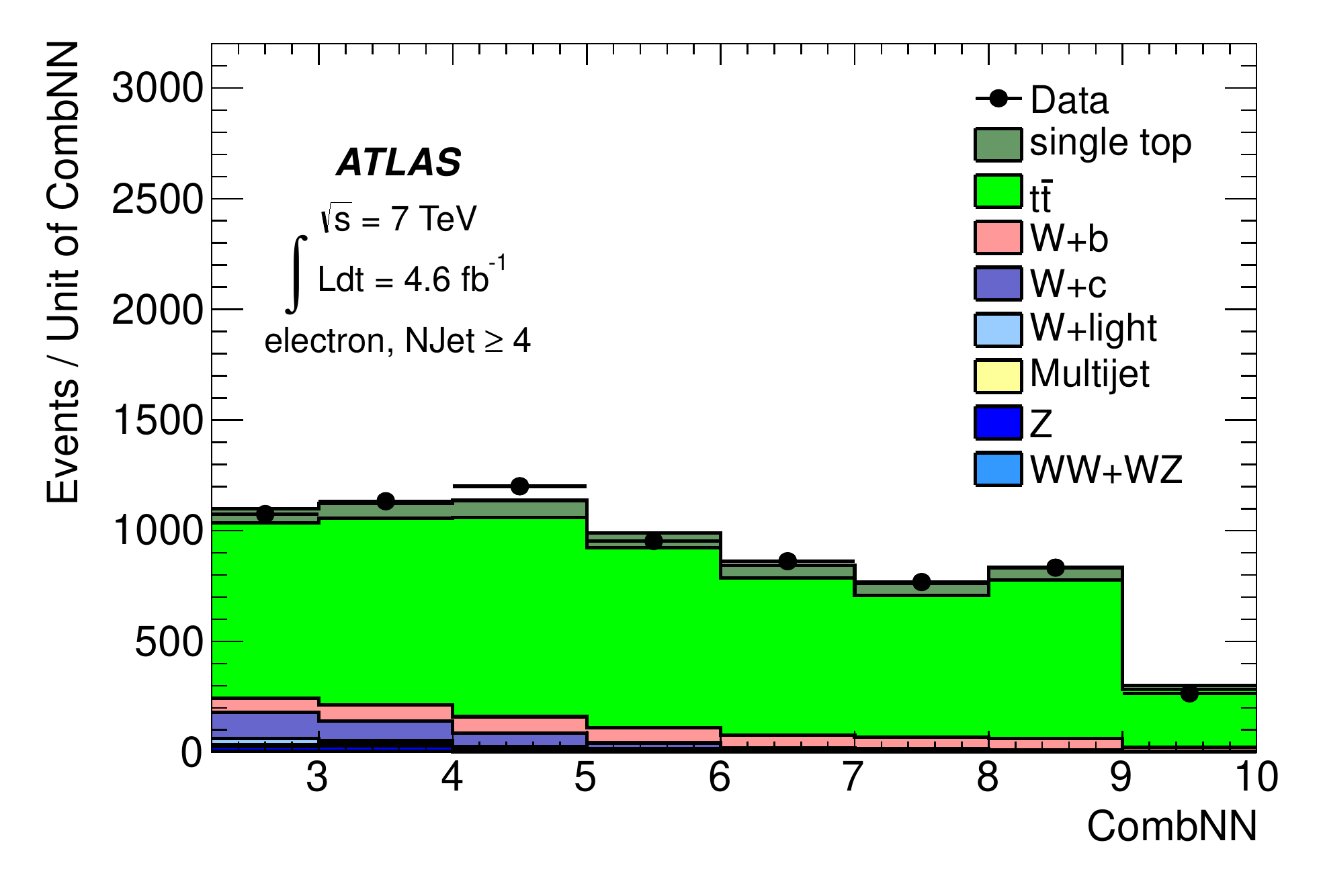}
\includegraphics[width=0.49\textwidth]{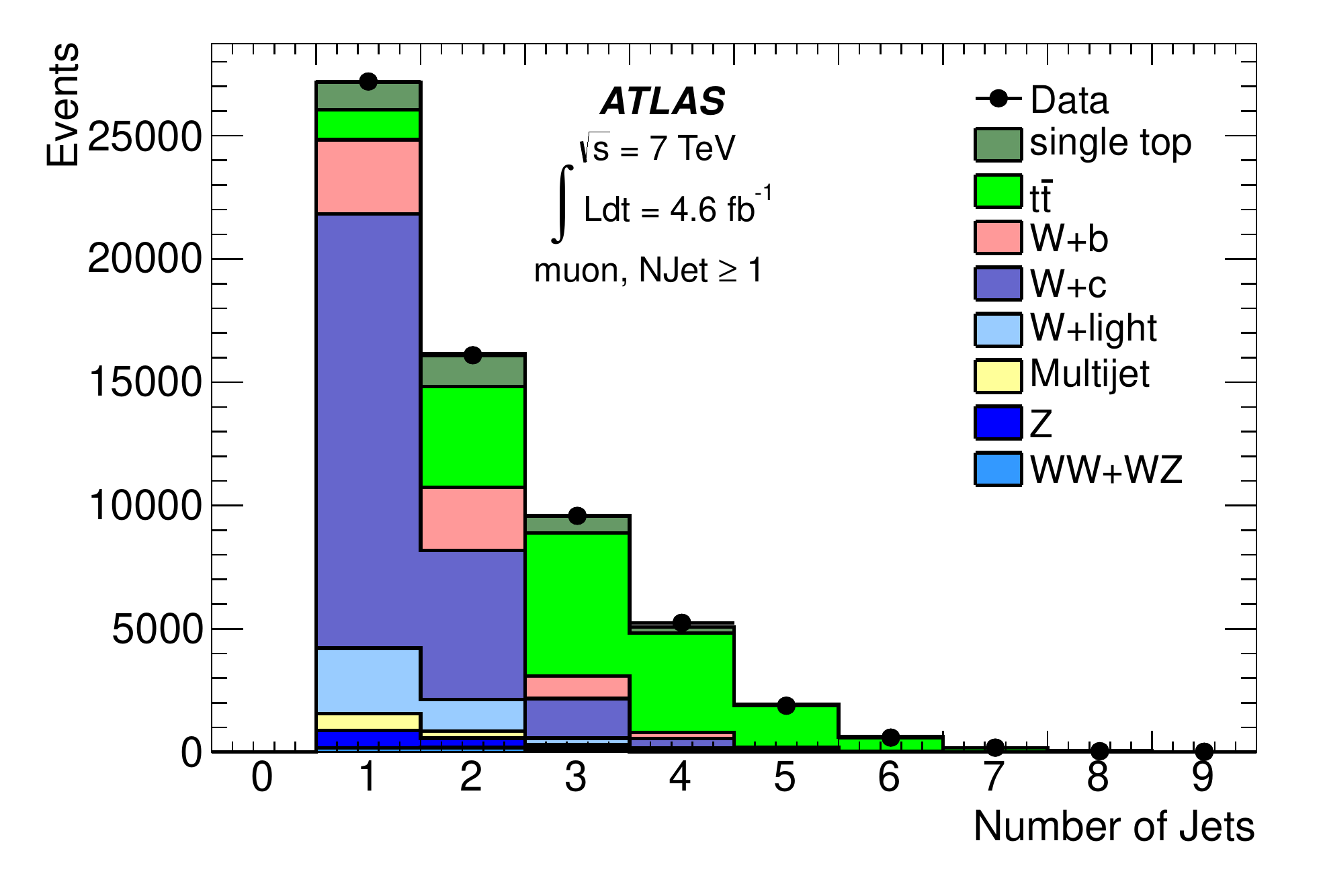}
\includegraphics[width=0.49\textwidth]{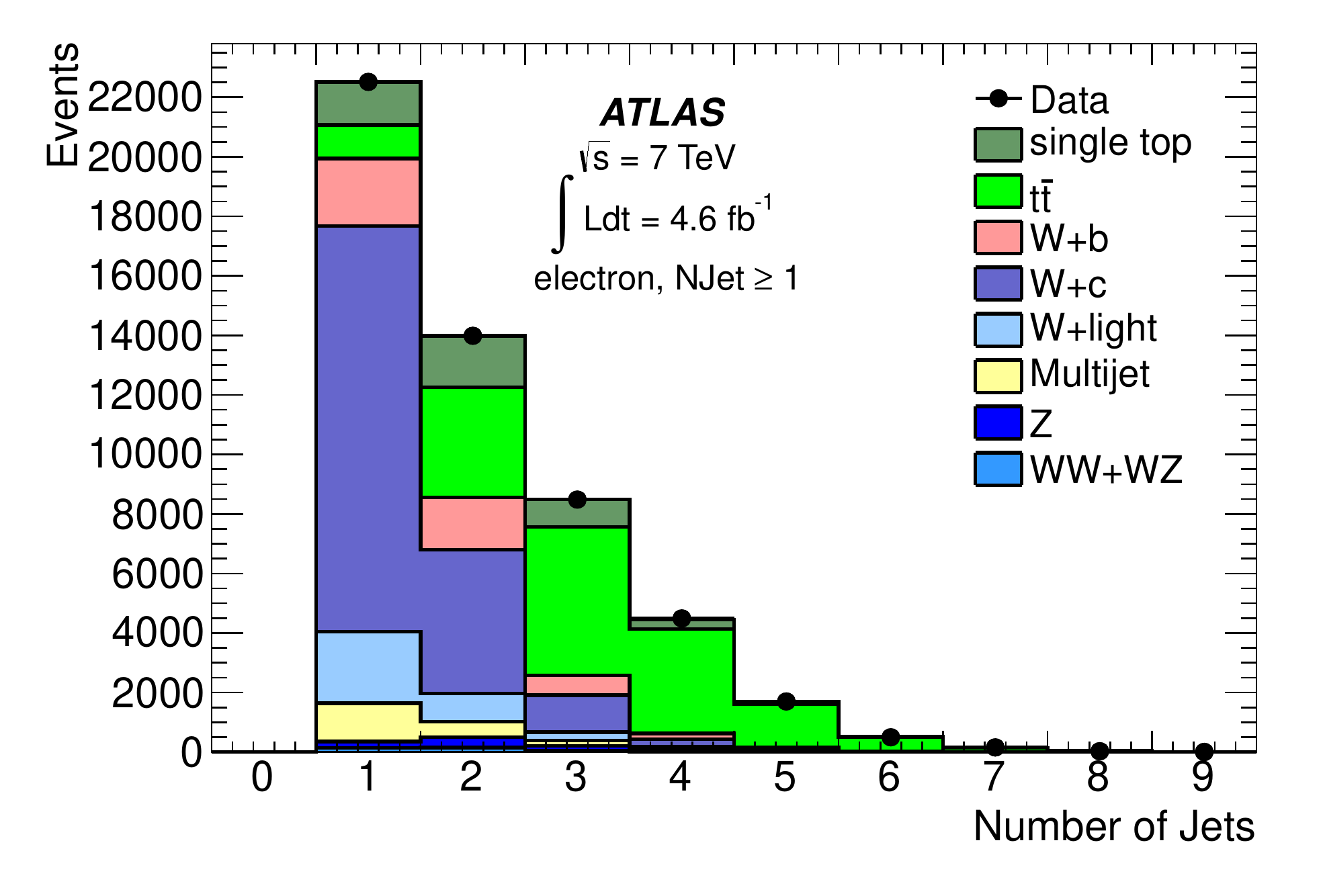}
\caption{ CombNN distributions for the $b$-tagged (CombNN$~ > 2.2$) jet in the ``\fourjet'' control region (top)  and number-of-jets distributions in the ``\onejetincl''  region (bottom) in data and MC simulation. The muon (electron) channel is shown on the left (right). MC samples are normalized to the results of the respective ML fits.}
\label{fig:ttcontrolfit}
\end{center}
\end{figure*}

\subsection{Single-top background}\label{sec:SingleTop_Background}

Single-top events containing a $W$ boson and at least one \bjet are, like \ttbar{} events,  an irreducible background for the \Wbjet{} signal. 
 In the \twojet region, where the single-top and \Wbjet contributions are comparable, kinematic observables are used to estimate the single-top normalization in data. 
The invariant mass of the combined $W$ boson and $b$-tagged jet system is computed for each event\footnote{The  $p_{z}$ of the neutrino is computed by setting the $W$ mass equal to the world average value of $80.399\GeV$~\cite{PDG}. In case of complex solutions, only the real part is considered.}, and the resulting $m(Wb)$ distribution, where single-top appears as a relatively narrow peak, is fitted. Example templates for the muon \twojet sample are shown in figure~\ref{fig:OverCombNN2}.

\begin{figure*}[h]
\begin{center}
\includegraphics[width=0.6\textwidth]{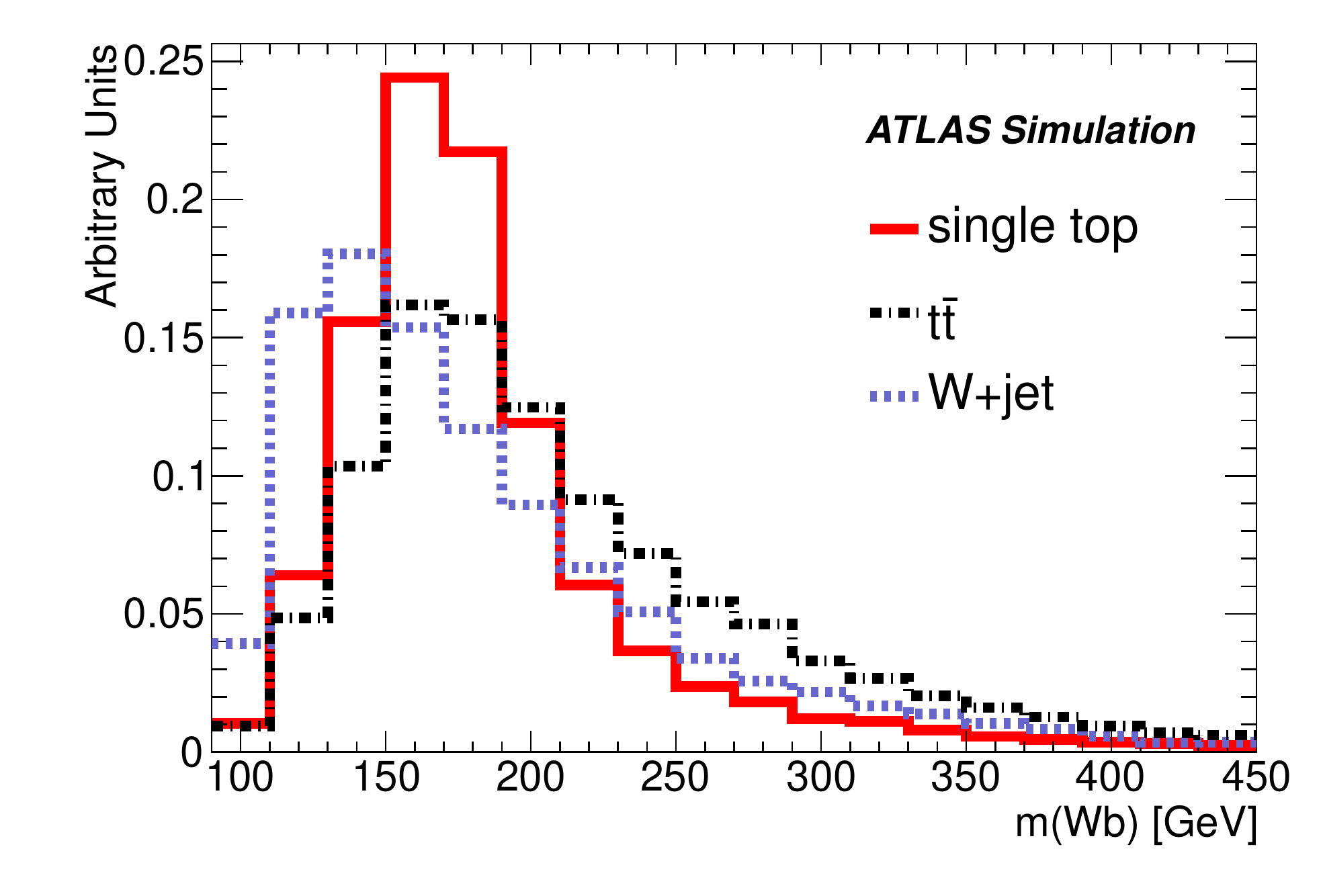}
\caption{Overlay of the $W$+jets, \ttbar\ and single-top $m(Wb)$  distributions  in MC simulation in the muon \twojet analysis region.}
\label{fig:OverCombNN2}
\end{center}
\end{figure*}

The $W$+jets and single-top template normalizations are free parameters of the fit. Gaussian constraints are applied to the multijet, \ttbar, $Z$+jets, and diboson backgrounds as described above. The single-top correction factors estimated by the fit in the electron and muon channel are, respectively, $1.13 \pm 0.15$ and $1.09 \pm 0.13$, and the corresponding fit projections are shown in figure~\ref{fig:stopcontrolfit}. 

These estimates of the single-top background contribution are verified with a  fit to the {\sloppy  $\HT= \pT^{\ell} + \MET + \sum\limits_{i=1}^{n} \pT^{\mathrm{jet}_i}$} distribution, where single-top events are expected, on average, to be harder than $W$+jets events. The corresponding single-top correction factors ($1.17 \pm 0.17$ and $1.12 \pm 0.11$ for the electron and muon channels, respectively)  are consistent with those obtained from the $m(Wb)$ fit. The $m(Wb)$-derived factors are therefore used to scale the \acermc\ single-top prediction in the \twojet region, and a 20\% uncertainty is assigned to its  normalization and applied as a Gaussian constraint in subsequent ML fits. 

In the \onejet region, where the expected single-top contribution is approximately half the size of the expected \Wbjet signal, the \acermc\ prediction is used, and a large normalization uncertainty (50\%) is assigned. This uncertainty reflects the difference observed in the single-top contribution estimated by the fit to the  $m(Wb)$ distribution and the fit to the \HT\ distribution in this analysis region.

\begin{figure*}[!ht]
\begin{center}
\includegraphics[width=0.49\textwidth]{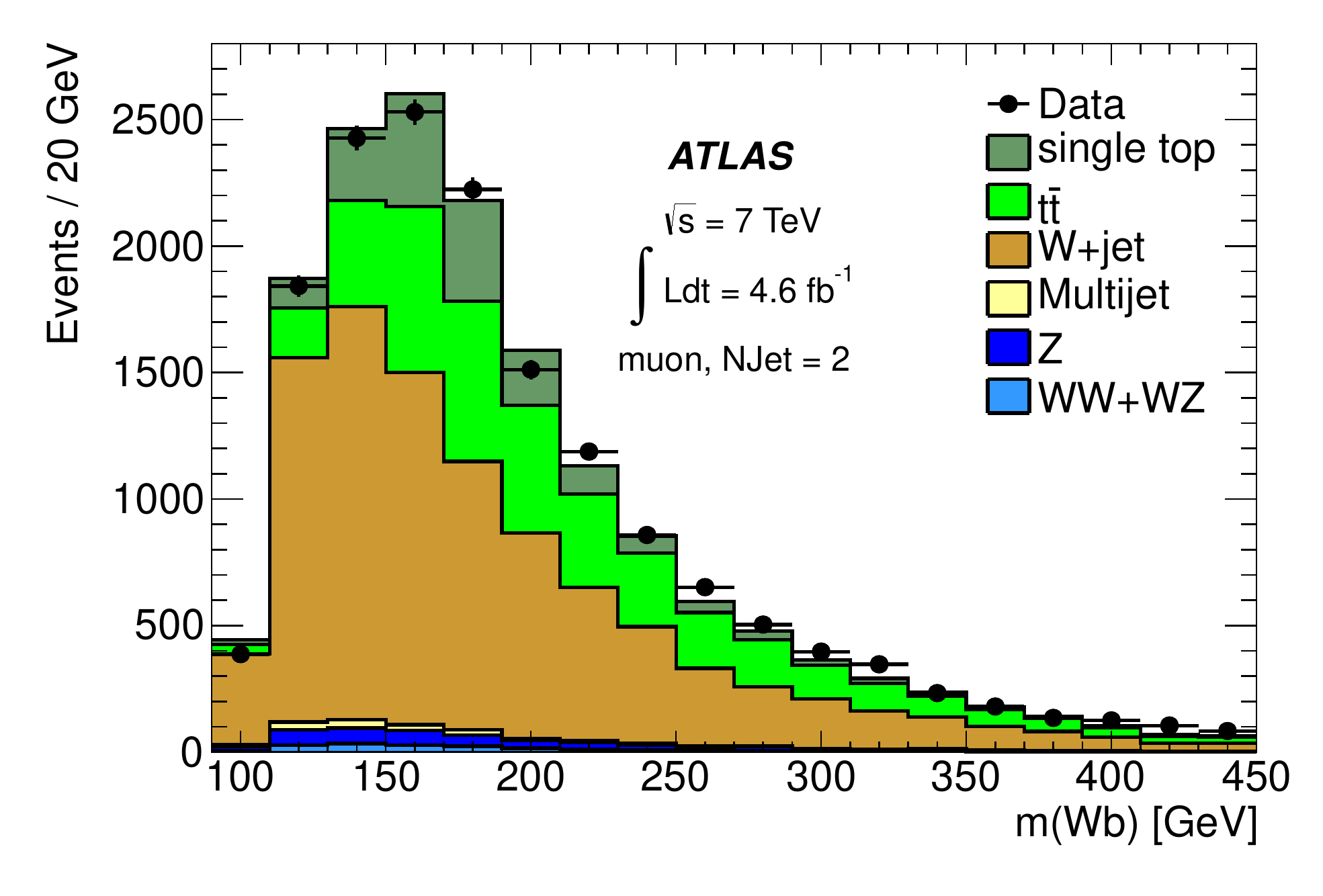}
\includegraphics[width=0.49\textwidth]{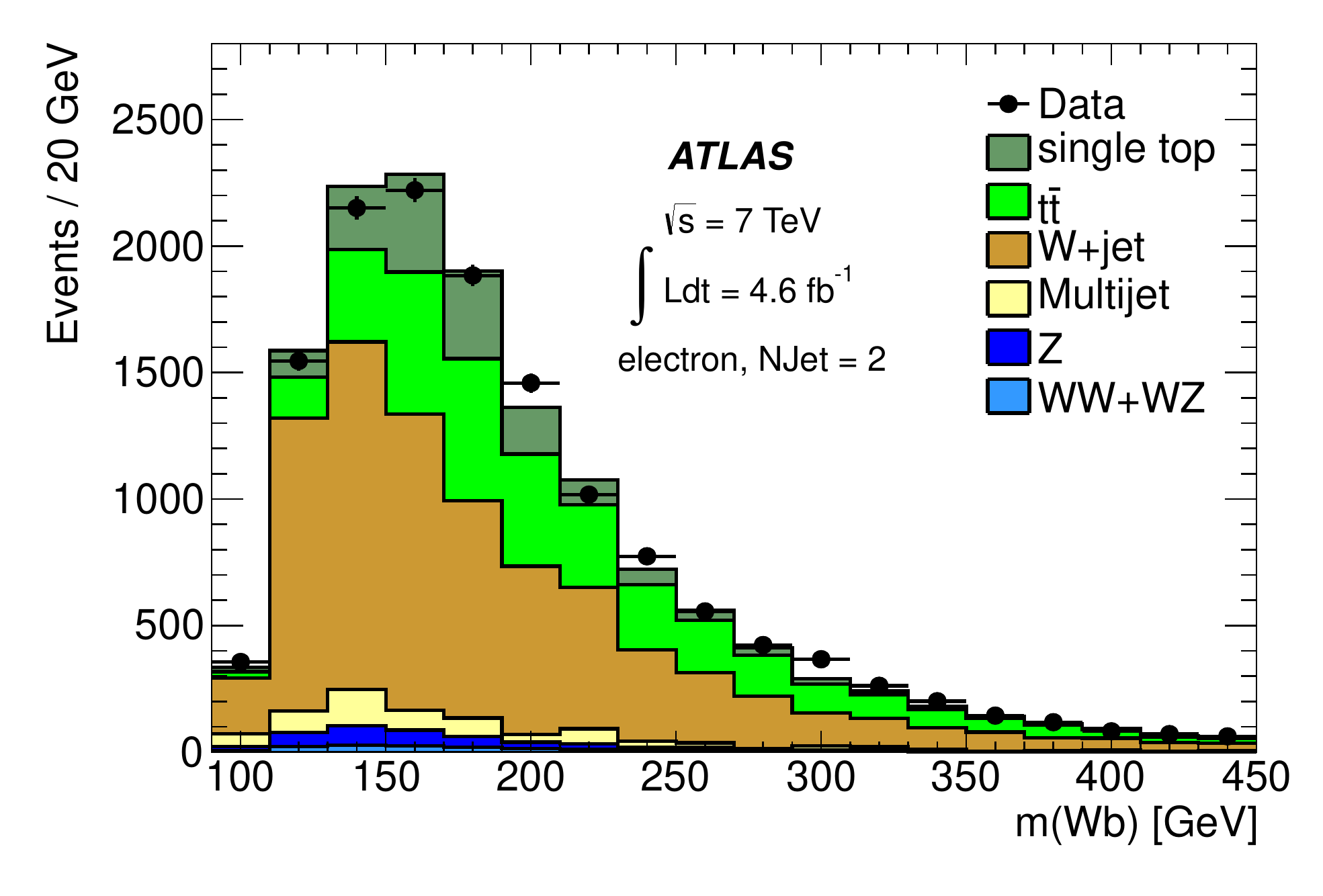}
\caption{Distributions of $m(Wb)$ in the \twojet region  in data and MC simulation for the muon (left) and electron (right) channels. The MC samples are normalized to the results of the respective ML fits. }
\label{fig:stopcontrolfit}
\end{center}
\end{figure*}

\subsection{$W$+jets backgrounds and signal}

The different response of $b$-jets, $c$-jets, and light-jets to the CombNN $b$-tagging algorithm is used to separate statistically the \Wbjet component from  the \Wcjet and \Wljet ones in each analysis region.
The \Wbjet,  \Wcjet and \Wljet normalizations are free parameters of the fit, while Gaussian constraints are applied to all other processes. 
In addition to the uncertainties discussed above for the multijet, \ttbar{} and single-top backgrounds, 10\% Gaussian constraints are assigned to the diboson and $Z$+jets backgrounds as discussed previously. 

The CombNN templates for the multijet component are extracted from data, while those from the other non-$W$ processes are extracted from the respective MC samples. For \Wbjet,  \Wcjet and \Wljet, the corresponding  templates are prepared in each analysis region using large \pythia-generated samples.  

The CombNN distributions normalized to the fit results are shown in figure~\ref{fig:fit}, and the number of \Wbjet and background events estimated by the fits, along with their statistical uncertainties, are summarized in table~\ref{tab:fitresults}. Table~\ref{tab:fitbeta} shows the correction factors estimated by the fit to the data compared to the prediction for each process.
While the electron and the muon sample backgrounds are treated as completely uncorrelated, the estimated background levels are found to be in good agreement across the 
channels. 
The behavior observed in the $Z$+jets background prediction in table~\ref{tab:fitresults}, when comparing the electron and muon channels in the 1-jet and 2-jet regions, is due to the different properties of $Z\to ee$ and $Z\to\mu\mu$ events in which one lepton is not reconstructed. In particular,  $Z\to ee$ events tend to have a higher number of jets (from the missing electron), while $Z\to\mu\mu$ events tend to have higher \MET\ (from the missing muon).
The \Wljet estimate does not
strongly affect the measured \Wbjet cross-section, as the \Wljet and
\Wbjet estimators have a 15\% correlation. Instead, the
\Wljet and \Wcjet estimators are 90\% anticorrelated, and a modest
change in the dominant  \Wcjet contribution can change the \Wljet
estimate significantly.

In the differential \pTb measurement, a separate fit to the CombNN distribution is performed in each analysis region in  four intervals of $b$-tagged jet \pt{}: 25--30~\GeV, 30--40~\GeV{}, 40--60~\GeV{} and 60--140 \GeV{}.  The background contributions are extrapolated to each \pt{} interval from the inclusive measurements, and the same Gaussian constraints as those of the inclusive fits are used. 
For the multijet background, this extrapolation is based on the $b$-tagged jet \pt\ spectrum found in the multijet templates extracted from data. For all other backgrounds, the extrapolation is based on Monte Carlo simulation.
\begin{figure*}[!ht]
\begin{center}
\includegraphics[width=0.49\textwidth]{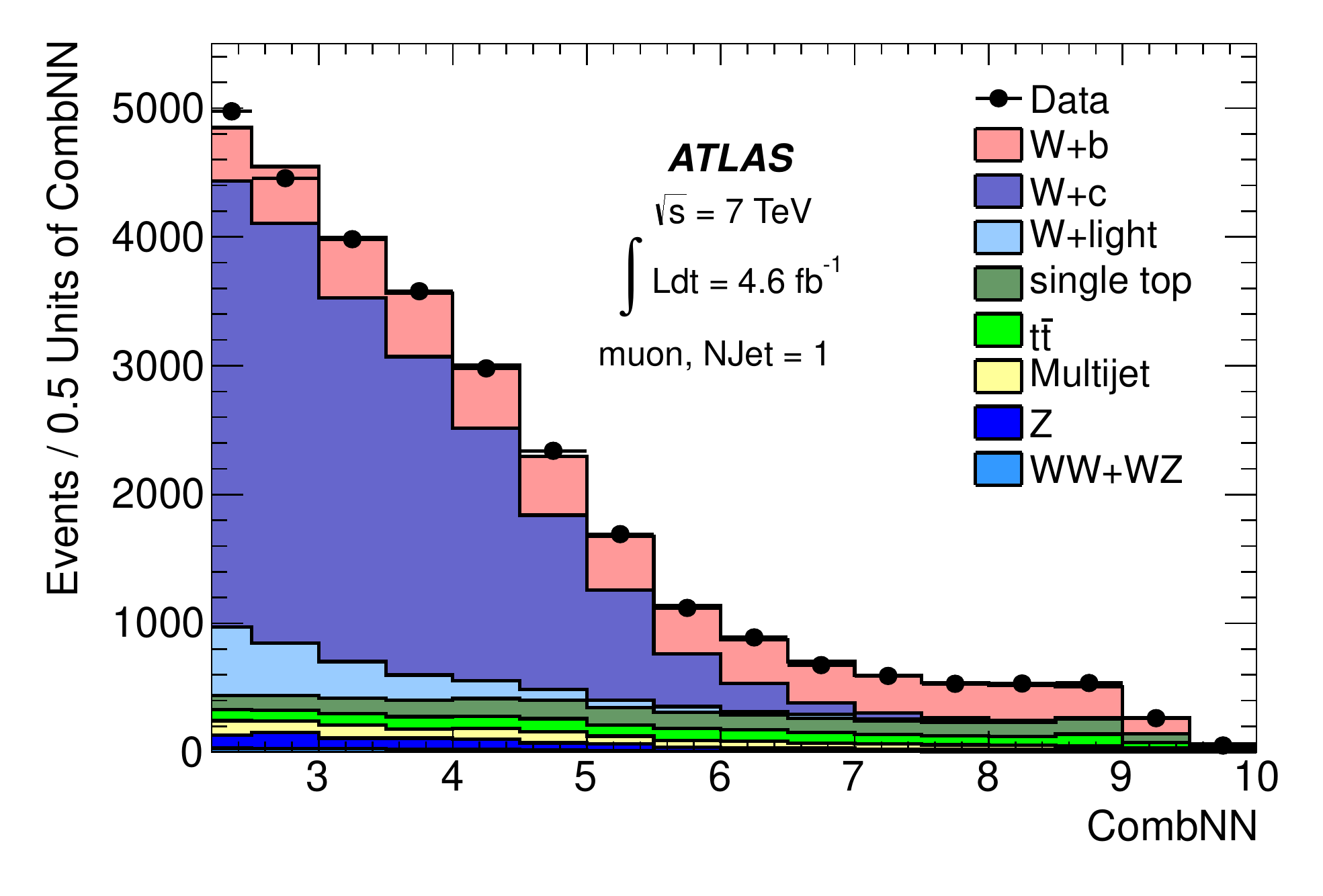}
\includegraphics[width=0.49\textwidth]{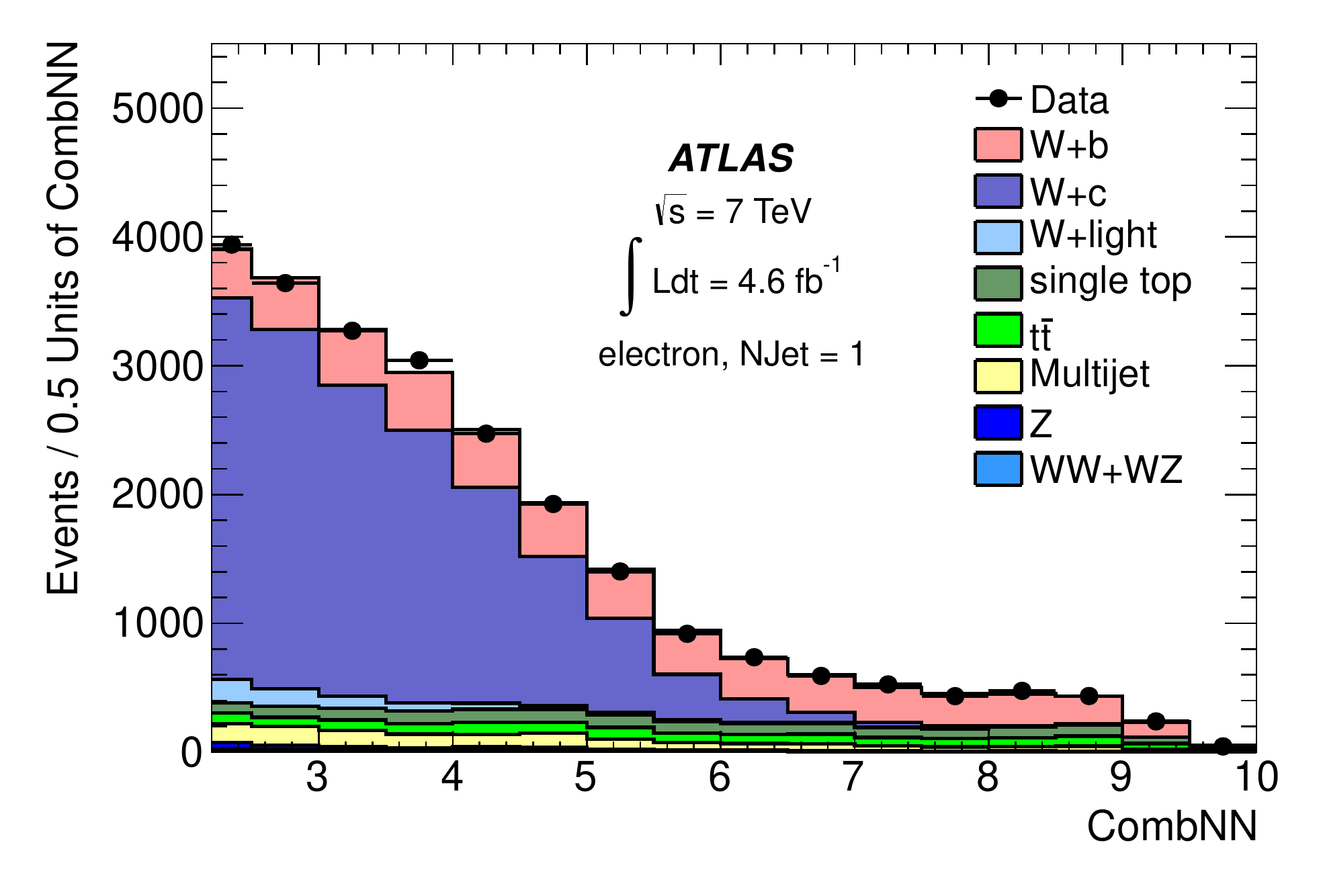}
\includegraphics[width=0.49\textwidth]{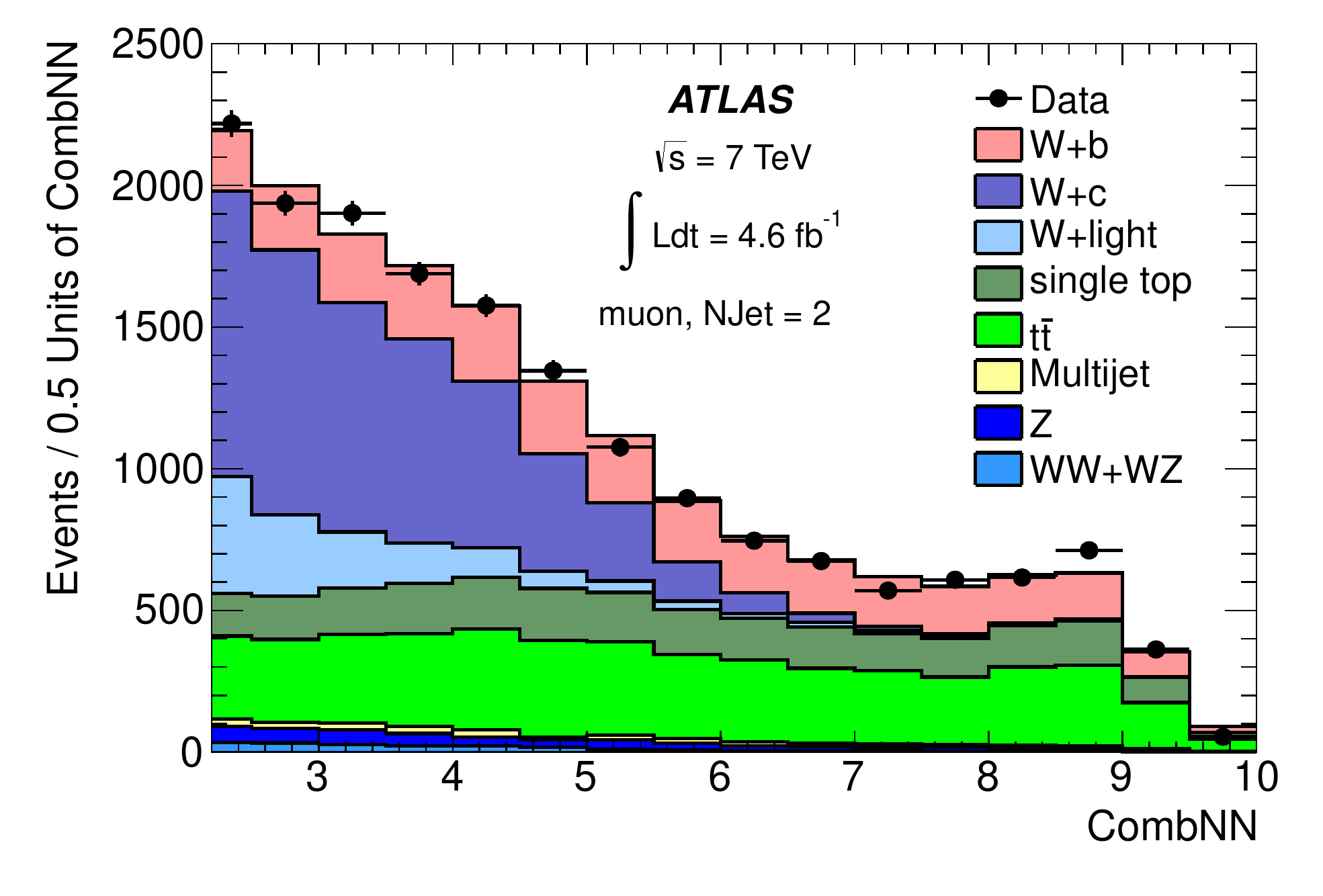}
\includegraphics[width=0.49\textwidth]{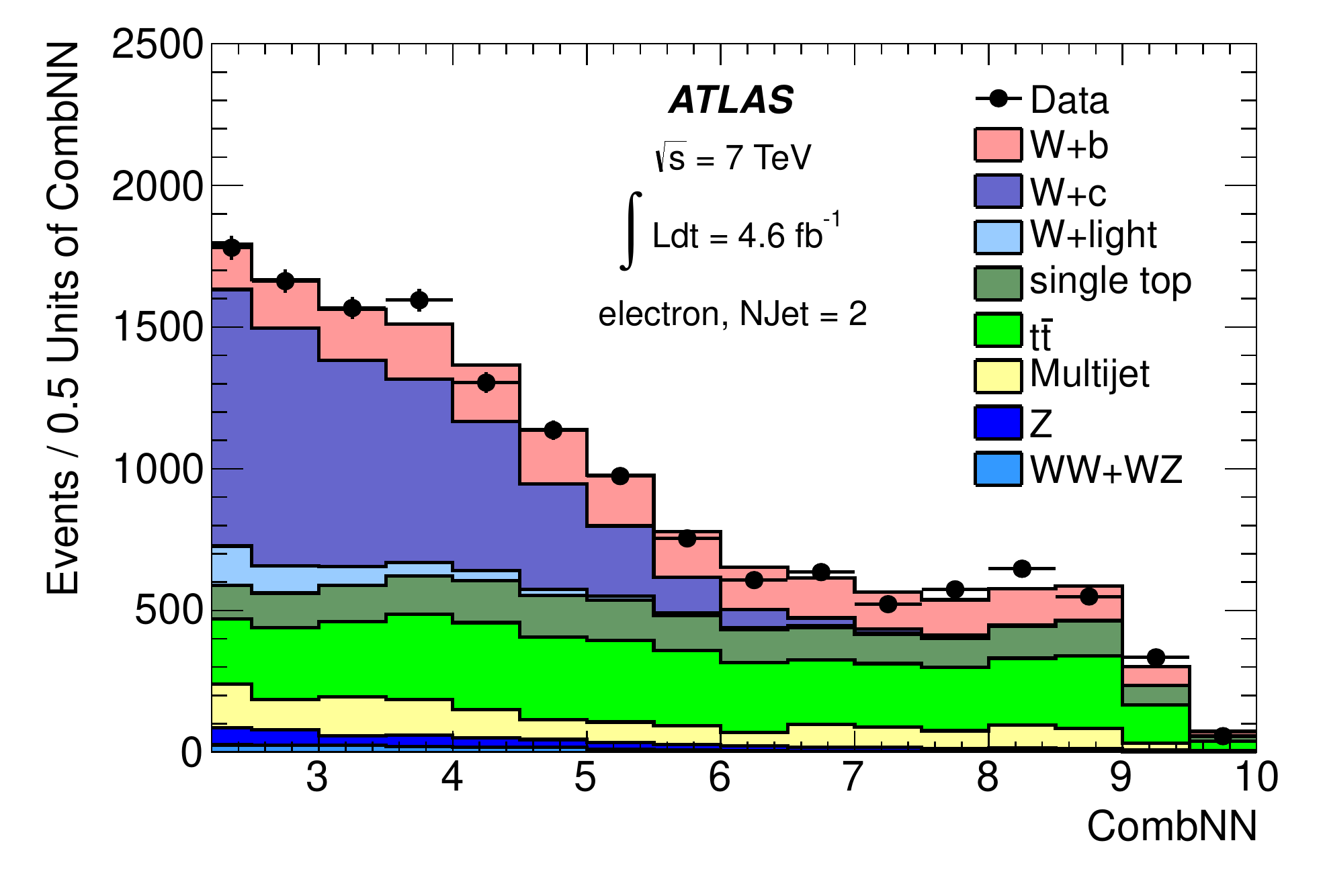}
\caption{CombNN distributions for the $b$-tagged jet in data and MC simulation, where the MC samples are normalized to the results of the ML fit, for the 1-jet (top) and 2-jet (bottom) analysis regions, in the muon (left) and electron (right) channels. }
\label{fig:fit}
\end{center}
\end{figure*}

\begin{table}[!htbp] 
\centering 
\begin{tabular}{c  | r @{~$\pm$~} l | r @{~$\pm$~} l | r @{~$\pm$~} l | r @{~$\pm$~} l }
\hline 
\hline
Process &  \multicolumn{2}{c|}{$\mu$ \onejet} & \multicolumn{2}{c|}{$e$ \onejet} & \multicolumn{2}{c|}{$\mu$ \twojet} & \multicolumn{2}{c}{$e$ \twojet} \\ 
\hline 
\hline 
 \Wbjet  &  ~5300 & 400 	& 4800 & 400  &  3000 & 260 & 2220 &  250  \\ 
\hline 
 \Wcjet  & 15600 & 600 	& 13300 & 500  &  4600 & 400 &  4000 & 400  \\ 
\hline 
\Wljet    & 1600 & 500  	& 500 & 500 	&  1170 & 330 & 490 & 320 \\ 
\hline 
\ttbar\          & 1230 & 120 & 1100 & 110  & 4300 & 400 & 3690 & 350  \\ 
\hline 
Single-top & 1700 & 500	 & 1400 & 500  & 2300 & 400 &  1810 & 350  \\ 
\hline 
Diboson & 181 & 18 	& 139 & 14   	&  185 & 18 &  155 & 15 \\ 
\hline 
$Z$+jets   & 770 & 70 	& 258 & 26  	&  397 & 40 & 365 & 37  \\ 
\hline 
Multijet    &   780 & 330		 & 1000 & 500  	 & 210 & 150 &  1220 & 290  \\ 
\hline 
\hline 
\end{tabular} 
\caption{Estimated event yields for the eight contributions to the four analysis regions, including the statistical uncertainty from the 
binned ML fit.}
\label{tab:fitresults} 
\end{table}

\begin{table}[!htbp] 
\centering 
\begin{tabular}{c | c | c | c | c }
\hline 
\hline
Process &  $\mu$ \onejet & $e$ \onejet & $\mu$ \twojet & $e$ \twojet \\ 
\hline 
\hline 
 \Wbjet  &  1.68 $\pm$ 0.14 & 1.98 $\pm$ 0.16  &  1.14 $\pm$ 0.10 & 1.16 $\pm$ 0.13  \\ 
\hline 
 \Wcjet  &  1.22 $\pm$ 0.04 & 1.30 $\pm$ 0.05  &  1.04 $\pm$ 0.09 &  1.10 $\pm$ 0.10  \\ 
\hline 
\Wljet    &  0.70 $\pm$ 0.22  & 0.28 $\pm$ 0.25 &  1.15 $\pm$ 0.33 & 0.67 $\pm$ 0.44 \\ 
\hline 
\ttbar\          &  1.00 $\pm$ 0.10 &1.00 $\pm$ 0.10  & 1.02 $\pm$ 0.10 & 1.01 $\pm$ 0.10  \\ 
\hline 
Single-top &  1.07 $\pm$ 0.34 & 1.02 $\pm$ 0.36  & 1.08 $\pm$ 0.19 &  1.01 $\pm$ 0.19  \\ 
\hline 
Diboson&   1.00 $\pm$ 0.10 & 1.00 $\pm$ 0.10   &  1.00 $\pm$ 0.10 &  1.00 $\pm$ 0.10 \\ 
\hline 
$Z$+jets   &   1.00 $\pm$ 0.10 & 1.00 $\pm$ 0.10  &  1.00 $\pm$ 0.10 & 1.00 $\pm$ 0.10 \\ 
\hline 
Multijet  &   1.12 $\pm$ 0.47 & 0.80 $\pm$ 0.40   & 0.67 $\pm$ 0.49 &  1.79 $\pm$ 0.42  \\ 
\hline 
\hline 
\end{tabular} 
\caption{Correction factors estimated by the binned ML fit to the CombNN distribution for each process in the four analysis regions, including the statistical uncertainty. 
The multijet, \ttbar\ and \twojet single-top factors are given with respect to their estimate in data.
The remaining factors are given with respect to the Monte Carlo expectations
normalized to the NLO (single-top, diboson) and inclusive NNLO ($W/Z$+jets) cross-sections.}
\label{tab:fitbeta} 
\end{table}

\section{Cross-section extraction}

The \Wbjet yields obtained from the CombNN fits are converted to a fiducial cross-section for \Wbjet times the branching ratio for each $W\rightarrow \ell \nu$ decay channel ($\ell = e,\mu$) using Monte Carlo simulation.
The unfolding procedure is defined with respect to the fiducial region introduced in table~\ref{tab:fiducialps}. It accounts for trigger and object reconstruction efficiencies (including the \bjet identification efficiency) after applying corrections for all known detector effects.
The small contribution (less than 5\%) from $W \rightarrow \tau \nu$,  where the $\tau$ decays to an electron or a muon, is not included in the fiducial region. 

The \alpgen\ Monte Carlo simulation is used to produce correction factors to account for two effects: events passing the fiducial selection which fail the reconstructed-level selection, and events which pass the reconstructed-level selection but originate from outside the fiducial region. 
These factors are applied to the inclusive \Wbjet yield in each analysis region to obtain a fiducial cross-section.

The differential \Wbjet cross-section is also extracted, in the \onejet and \twojet regions, as a function of the transverse momentum of the leading $b$-jet, \pTb, using the same bins as the CombNN differential fits.
The measured quantity is therefore d$\sigma_{\mathrm{fiducial}}$/d\pTb. 
For this measurement, the correction factors mentioned above are produced in each \pTb bin.
\alpgen\  Monte Carlo events which pass both  reconstruction and fiducial selections are used to generate a response matrix to account for bin-to-bin migration effects between the reconstructed and generator-level distributions of \pTb.
This response matrix is applied through an iterative Bayesian technique~\cite{iterDAgostini}, in which the MC prediction is used as the initial prior, and three successive iterations are performed to remove the bias from the initial distribution.

The stability of the unfolding procedure is tested by comparing the unfolded spectra after three iterations with  those obtained using two and four iterations, yielding consistent results. 
The bias introduced by the choice of prior is tested by creating an alternative sample, and unfolding it using the nominal response matrix; after three iterations, the unfolded distribution is significantly different from the nominal initial prior, and reproduces correctly the alternative generated distribution.

Different jet bins and lepton flavour channels are combined to yield more precise measurements of the \Wbjet cross-section.
In order not to introduce new assumptions on the background normalizations, the \Wbjet  yields are added after the CombNN fit, and their sum is unfolded using correction factors and response matrices obtained from Monte Carlo simulated events in the combined channels. This procedure is performed for each systematic variation and, in order to take into account the correlation of systematic uncertainties, correlated  uncertainties are varied simultaneously in the samples being combined.

\section{Systematic uncertainties}

Several sources of systematic uncertainties on the measured \Wbjet cross-section are considered. Each source may affect the background estimation in the control regions, the results of the CombNN fits, and the unfolding factors and response matrices. The strategy described here is used in all the jet multiplicity regions and \bjet \pT\ intervals.

The effect of each systematic source on the estimated number of \Wbjet events is quantified using pseudo-experiments. For a given systematic variation, new sets of signal and background templates are prepared which may differ  in both shape and normalization from the reference set used in the fit to data. The modified templates are used to generate pseudo-data samples that are fitted using the reference templates. In these pseudo-experiments, the same background constraints as those used in the fit to data are applied. Finally, the quoted fractional systematic uncertainty associated with a given source is defined as the ratio $ (\bar{N}_{\rm fit} - \mu_{\rm gen})~/~\mu_{\rm gen}$, where $ \bar{N}_{\rm fit} $ is the mean of the estimator of the number of \Wbjet events  and $\mu_{\rm gen}$ is the number of \Wbjet events used in the pseudo-data generation for that particular systematic source.

The full analysis procedure is repeated for each systematic variation:  the multijet, $t\bar{t}$ and single-top contributions are estimated in the corresponding control sample or distribution, and the CombNN distribution is fitted in the analysis regions after propagating the new background estimates. 

Systematic uncertainties in the unfolding process are accounted for by using each systematically varied signal Monte Carlo sample to generate an alternative response matrix and set of unfolding factors. The difference in fiducial cross-section obtained when using the alternative Monte Carlo in place of the default one is quoted as the systematic uncertainty in the measurement. For systematic uncertainties that are split into an upwards and downwards variation, the unfolding is performed twice, and only the largest of the two resulting variations is taken as a symmetric uncertainty.

Most of the systematic effects considered here influence both the fitting and unfolding steps. In these cases, the systematic effects are propagated coherently and for a given systematic source the corresponding estimated \Wbjet yields are unfolded using the corresponding response matrix. 

Background normalizations (multijet, \ttbar{}, single-top, diboson and $Z$+jets)  are treated as nuisance parameters  of the ML fit to the CombNN distribution. As such, background normalization uncertainties are accounted for in the uncertainty on the number of \Wbjet events estimated by the fit. In the unfolding,  the statistical uncertainty of the fiducial cross-section is evaluated using pseudo-experiments based on the uncertainty on the number of \Wbjet events estimated by the fit. For reference, if the background normalizations are fixed and the corresponding nuisance parameters are removed, the uncertainty on the number of \Wbjet events decreases by almost a factor of two.

The following effects are found to be non-negligible for the cross-section measurements:

{\bf Jet energy scale and resolution}. The uncertainty on the jet energy scale (JES) is derived from data and from Monte Carlo simulation~\cite{ref:JES, jeseta, jes, JESpileup, JESInSituZ, JESInSituGamma}, and varies between 3\% and 14\% depending on the jet \pT\ and pseudorapidity. This uncertainty includes effects arising from the dependence of the jet response on the pile-up. It also accounts for differences between the calorimeter responses to light-quark-, gluon-, and heavy-quark-initiated jets, and for additional low-momentum jets found within $\Delta R = 0.8$ of each jet considered.
Uncertainties related to the jet energy resolution (JER) are derived from the jet response asymmetry measured in dijet events in data~\cite{ref:JES, jes}. 

The effects of the JES and JER uncertainties are quantified using alternative signal and background Monte Carlo templates in which the jet energy is modified by $\pm 1 \sigma$ or smeared, respectively. They represent the dominant sources of systematic uncertainties on the measured  \Wbjet fiducial cross-sections and are found to be in the range 10--50\%, depending on the jet multiplicity and \pT\ interval considered. 

{\bf Initial-state and final-state radiation (ISR/FSR)}.
Uncertainties on ISR and FSR affect the extrapolation of the \ttbar\ contribution in the analysis regions, as well the single-top Monte Carlo expectation in the \onejet region and the data-driven single-top estimate in the \twojet region. These effects are evaluated using the \acermc\ generator interfaced to \pythia{}, and by varying the parameters controlling ISR and FSR in a range consistent with experimental data~\cite{isrfsr1}. Their effect on the final cross-section measurements depends strongly on the jet \pT\ interval and varies between 2\% and 30\%.

{\bf \emph{b}-tagging efficiency calibration}. The calibration of the $b$-tagging efficiency is performed using control samples in data~\cite{ref:BtaggingSF}. Uncertainties on these calibrations are estimated separately  for light-jets, $c$-jets and $b$-jets as a function of the \pT\ and $\eta$ of the jet~\cite{ref:BtaggingSFC, ref:BtaggingSFL}. These uncertainties affect both the \Wbjet selection efficiency and the shape of the CombNN templates. The corresponding impact on the measured cross-section is estimated independently for $b$-jets, $c$-jets and light-jets to be in the range 1--8\%. 

{\bf Monte Carlo modelling}. The uncertainty related to the \alpgen\ \Wbjet fiducial acceptance modelling is estimated using alternative \Wbjet samples, generated using different settings. Specifically,  the functional form of the factorization scale is varied; the set of parton distribution functions (PDF) is changed from CTEQ6L1~\cite{CTEQ6L1} to MRST2002LO~\cite{MRST2002}; the renormalization and factorization scales are halved and doubled; finally, the minimum jet \pT\ used in the MLM matching is decreased (increased)  to 15 (25) GeV, from the reference value of 20 GeV. 

Because of the mild dependence of the CombNN template shape on the jet \pT, an additional systematic uncertainty due to the $c$-jets and $b$-jets \pT\ modelling is quoted. This uncertainty is estimated using  the full \herwig\ parton shower sample of $c$-jets and $b$-jets in place of the \alpgen\  matrix element ones. The full \herwig\ parton shower $c$-jets and $b$-jets spectra are found to be softer than the corresponding spectra produced by the \alpgen\  matrix element and the difference between the two is larger than any differences observed with the  alternative \alpgen\ samples mentioned above.

The systematic uncertainties on the measured cross-section related to the Monte Carlo modelling are in the range 2--8\%.

{\bf Topological cluster energy scale and pile-up modelling}.  These account for the contribution to the {\MET} uncertainty due to uncertainties on the energy measurement of low-momentum jets and calorimeter cells that are not associated with electrons, muons or jets, as well as the uncertainty on the modelling of pile-up~\cite{METuncert}. Their effect is estimated to be in the range 2--6\%.

{\bf CombNN weight templates shape}. Uncertainties on the CombNN shape of $b$-jets, $c$-jets and light-jets are quantified independently. These uncertainties affect the measured cross-section by changing the results of the CombNN fits, but they do not affect the unfolding process.

The systematic uncertainty associated with the \bjet CombNN shape is estimated using data. Events with at least four jets, two of which must be $b$-tagged, are selected. These events form a sample of \ttbar~candidates whose leading jet is a real \bjet approximately $95\%$ of the time, as estimated in Monte Carlo simulation. This clean sample of \bjet candidates is used to compute the ratio of the corresponding CombNN distribution in data to that in MC simulation. This ratio is applied to all the \bjet templates used in the CombNN fit and the new set of templates is used to assess the corresponding systematic uncertainty.

The systematic uncertainty associated with the \cjet CombNN shape is estimated by preparing alternative CombNN templates for \cjets. Both the \cjet and \bjet CombNN shapes are sensitive to the description of the $b$-hadron and $c$-hadron branching ratios and in particular to the number of charged particles produced in their decay vertices. While it is possible to define a clean and almost unbiassed control sample of \bjets in the data to check any possible discrepancy with the Monte Carlo simulation, the same is not feasible for \cjets. The alternative $c$-jet templates are then obtained by varying artificially the relative contribution of events with different track multiplicity associated with secondary vertices. Variations of 10\% are considered in each track multiplicity bin, consistent with the relevant uncertainties in the $c$-hadron branching ratios.

Finally, a light-jet shape extracted from \herwig\ is used in place of the one obtained from \pythia\ to assess the systematic uncertainties on the light-jet CombNN shape. The effects of the variations of \bjet and \cjet CombNN shapes on the final cross-section measurements range between 2\% and 8\%, while the effect of the light-jet CombNN shape variation is negligible. 

{\bf Multijet background CombNN shape}.
The systematic uncertainty on the multijet template shape is assessed using a control region defined by $\MET < 25 \GeV$ and \WmT$ <40\GeV$. 
Any mismodelling observed in this region is transported to the signal region and used to generate alternative multijet shapes, both in the electron and muon samples. The corresponding effect on the measured cross-section is larger in the electron sample where it is in the range 1--10\% depending on the jet multiplicity and \pT\ interval considered.

{\bf Others}.
Uncertainties related to the lepton trigger and reconstruction efficiencies are evaluated using tag-and-probe measurements in $Z\rightarrow \mu\mu$ and $Z\rightarrow ee$ events~\cite{TagProbeMu,TagProbeEle}. Similarly, 
the $Z$-mass peak is used to determine the lepton momentum scales and resolutions and the corresponding uncertainties~\cite{MomentumScaleMu,TagProbeEle}. The effect of these sources of uncertainties on the \Wbjet cross-section is between 1\% and 2\%.
A 3.9\% uncertainty on the integrated luminosity is also included~\cite{bib:Lum2}.

\section{Results}

The unfolded result for the fiducial \Wbjet cross-section is presented in figure~\ref{Fig:resultsinclusive}, while the measured differential d$\sigma$/d\pTb distributions are shown in figure~\ref{Fig:Unfold:DataPtB1all}. The numerical values corresponding to the combination of the electron and muon channels are shown in tables~\ref{tabinclusive}--\ref{tabdiff2}, where details of the systematic uncertainties and correlation matrices for the statistical and systematic uncertainties are also presented. The measured cross-sections for the 1-jet, 2-jet and 1+2-jet fiducial regions are:

$\sigma_\mathrm{fid}$ (1 jet) = $5.0 \pm 0.5~\mathrm{(stat)}\pm 1.2~\mathrm{(syst)~pb}$, 

$\sigma_\mathrm{fid}$ (2 jet) = $2.2 \pm 0.2~\mathrm{(stat)}  \pm 0.5~\mathrm{(syst)~pb}$, 

$\sigma_\mathrm{fid}$ (1+2 jet) = $7.1\pm 0.5~\mathrm{(stat)} \pm 1.4~\mathrm{(syst)~pb}$.

The results are compared to the NLO predictions of \mcfm\ and \powheg{}, and to the \alpgen\ predictions scaled by the NNLO normalization factor for the inclusive $W$ cross-section~\cite{FEWZ}. 
Both the \alpgen\ and \powheg\ predictions implement a 4-flavour number scheme (4FNS) calculation, while the \mcfm\ prediction,  following the calculation described in ref.~\cite{ref:4FNS5FNS}, includes terms which use the 5-flavour number scheme (5FNS) to account for the presence of $b$-quarks in the initial state originating from parton distribution functions. 

\begin{figure*}[hb]
\begin{center}
\includegraphics[width=0.7\textwidth]{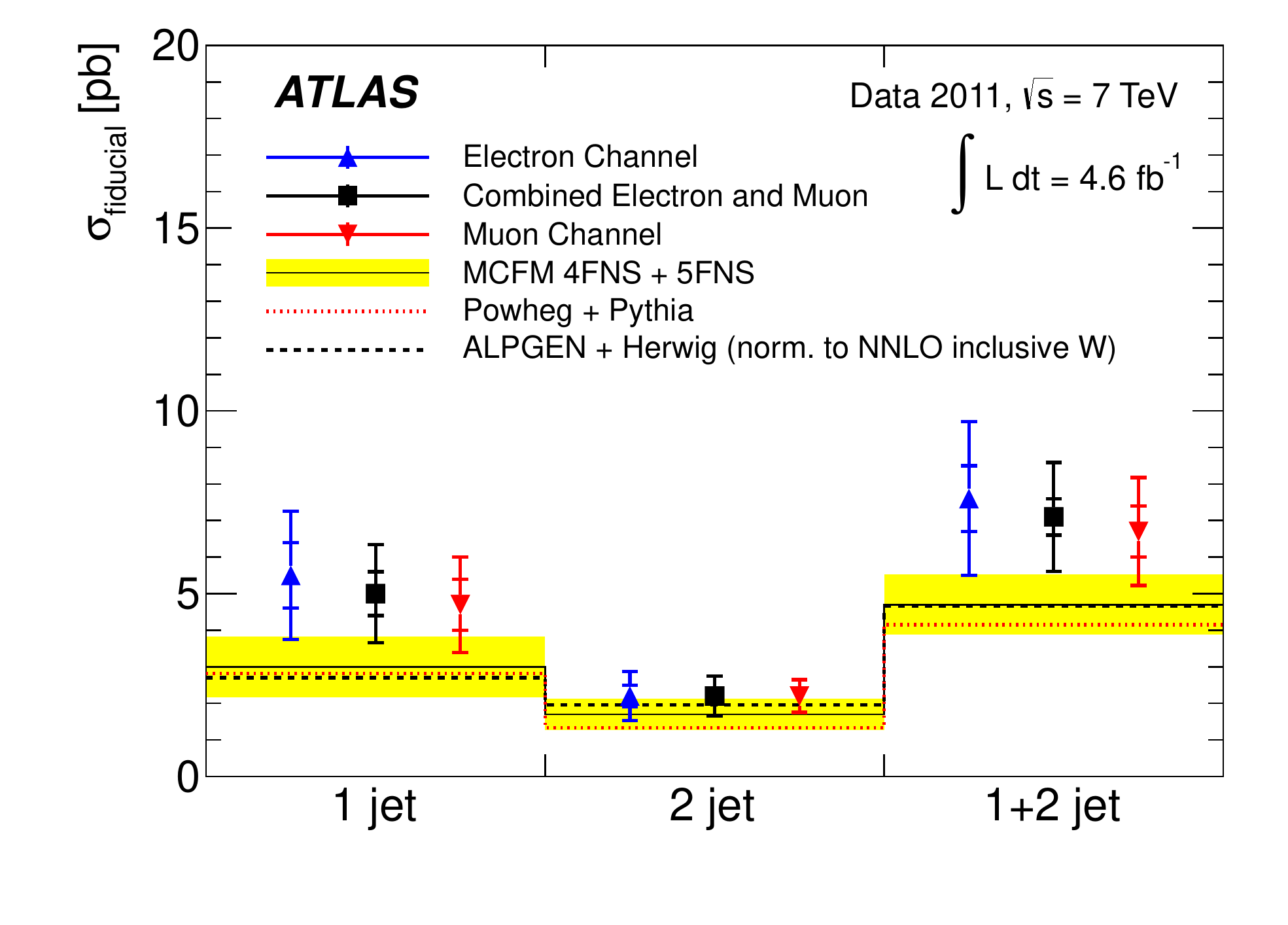}
\caption{Measured fiducial cross-sections with the statistical (inner
    error bar) and statistical plus systematic (outer error bar)
    uncertainties in the electron, muon, and combined electron and muon
    channels.  The cross-sections are given in the 1-jet, 2-jet, and 1+2-jet
    fiducial regions.  The measurements are compared with
    NLO predictions calculated with \mcfm~\cite{ref:4FNS5FNS} and corrected for hadronization 
   and double-parton interaction (DPI) effects.  The yellow bands
    represent the total uncertainty on the prediction. It is obtained by
    combining in quadrature the uncertainties resulting from variations 
    of the renormalization and factorization
    scales, the PDF set, the DPI model and non-perturbative corrections.
    The NLO prediction from \powheg\ interfaced to \pythia, 
    corrected for DPI effects, and the
   prediction from \textsc{Alpgen} interfaced to
    \textsc{Herwig} and \textsc{Jimmy} and scaled by the NNLO inclusive $W$ normalization factor are also shown.  }
 \label{Fig:resultsinclusive}
\end{center}
\end{figure*}

\begin{figure}[!h]
    \begin{center}
  \includegraphics[width=0.49  \textwidth]{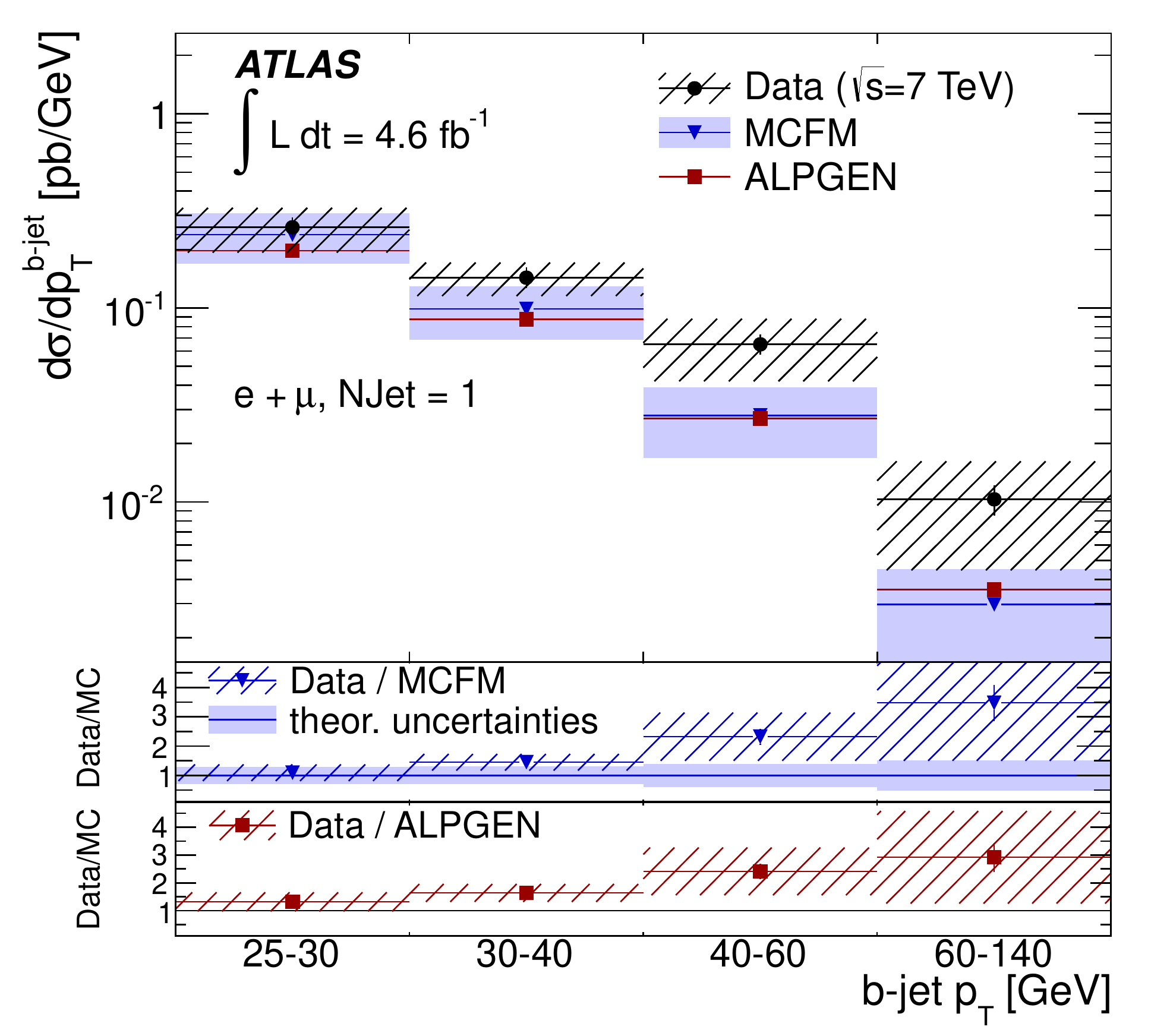}
    \includegraphics[width=0.49  \textwidth]{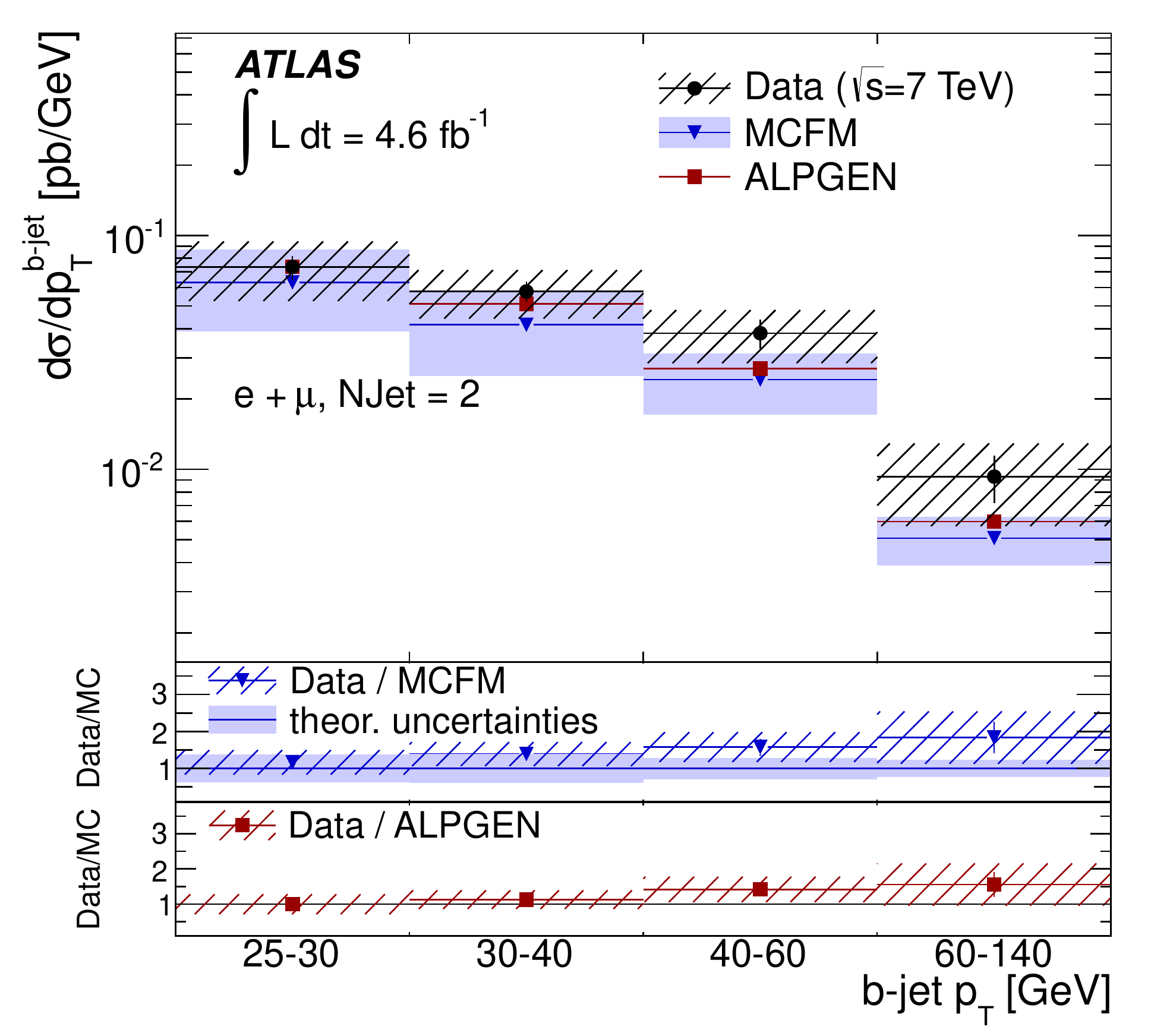}
    \caption{Measured differential \Wbjet cross-sections with the statistical plus systematic uncertainties as a function of \pTb in the 1-jet (left) and 2-jet (right) fiducial regions, obtained by combining the muon and electron channel results. 
        The measurements are compared to the \mcfm\ predictions and to the \alpgen{} predictions interfaced to \herwig{} and \jimmy\ and scaled by the NNLO inclusive $W$ normalization factor. The ratios between measured and predicted cross-sections are also shown.}
        \label{Fig:Unfold:DataPtB1all}
    \end{center}
\end{figure}

\begin{table*}[ht]
\caption{Measured fiducial \Wbjets cross-sections 
  for the combination of the electron and muon channels with statistical and systematic uncertainties
  and breakdown of relative systematic uncertainties per jet multiplicity, and combined across jet bins. }
\begin{center}
\begin{tabular}{l c c c}
\hline\hline
 \multicolumn{4}{c}{Fiducial cross-section [pb]} \\
\hline
& 1 jet & 2 jet &1+2 jet\\
\hline
     $\sigma_\mathrm{fid}$ &          $5.0$ &           $2.2$ &         $7.1$ \\
  Statistical uncertainty&            $0.5$ &        $0.2$ &         $0.5$ \\ 
   Systematic uncertainty&        $1.2$ &        $0.5$ &      $1.4$ \\ 
\hline
\hline
 \multicolumn{4}{c}{Breakdown of systematic uncertainty [\%]} \\
\hline
Jet energy scale &     $15 $ &       $15 $ &     $15 $ \\ 
Jet energy resolution  &      $14 $ &       $4 $ &       $8 $ \\ 
\bjet efficiency 		&        $6 $ &        $4 $ &        $5 $ \\ 
$c$-jet efficiency	&        $1 $ &        $1 $ &        $0 $ \\ 
light-jet efficiency	&        $1 $ &        $3 $ &        $2 $ \\ 
ISR/FSR 		&        $4 $ &        $8 $ &        $3 $ \\ 
MC modelling	&        $8 $ &        $4 $ &        $6 $ \\ 
Lepton resolution 	&        $1 $ &        $1 $ &        $0 $ \\ 
Trigger efficiency	&        $1 $ &        $2 $ &        $2 $ \\ 
Lepton efficiency	&        $1 $ &        $2 $ &        $1 $ \\ 
\MET\  scale 	&        $3 $ &        $6 $ &        $2 $ \\ 
\MET\  pile-up 	&        $2 $ &        $2 $ &        $2 $ \\ 
\bjet  template &       $3 $ &         $5 $ &        $4 $ \\ 
$c$-jet  template &    $4 $ &        $2 $ &        $3 $ \\ 
light-jet  template &  $0 $ &        $0 $ &         $0 $ \\ 
Multijet template &    $2 $ &         $2 $ &        $2 $ \\ 
\hline
Total syst. uncertainty 	&	$24 $ &       $23 $ &     $20 $ \\ 
\hline\hline
\end{tabular}
\label{tabinclusive}
\end{center}
\end{table*}

\begin{table*}[ht]
\caption{Measured fiducial \Wbjets cross-section in the 1-jet  region with statistical and systematic uncertainties
and their correlations in bins of \pTb.}
\begin{center}
\begin{tabular}{l c c c c}
\hline\hline
 \multicolumn{4}{c}{Fiducial cross-section, 1 jet  } \\
\hline\hline
\pTb [\GeV{}] &  [25, 30]         & [30, 40]         & [40, 60]         & [60, 140]       \\
\hline
d$\sigma$/d\pTb  [nb/\GeV{}]               & 259  & 143  & 65  & 10.3  \\
\hline
 Statistical Uncertainty  (\%)                & 9   & 6   & 12  & 18  \\
 Systematic Uncertainty  (\%)                & 24 & 19   & 33  & 54  \\
\hline
 Correlation coefficients of   & 1      & 0.415  & $- 0.38$  & $- 0.02$  \\
statistical uncertainties &  & 1      & $-0.01$  & $ -0.17$  \\
&  &  & 1      & $-0.14$  \\
&  &  &  & 1      \\
\hline
Correlation coefficients of &  1      & 0.893  & 0.740  & 0.582  \\
systematic uncertainties &  & 1      & 0.887  & 0.750  \\
 &  &  & 1      & 0.875  \\
 &  &  &  & 1      \\
\hline\hline
\end{tabular}
\label{tabdiff1}
\end{center}
\end{table*}

\begin{table*}[ht]
\caption{Measured fiducial \Wbjets cross-section in the 2-jet  region with statistical and systematic uncertainties
and their correlations in bins of \pTb.}
\begin{center}
\begin{tabular}{l c c c c}
\hline\hline
 \multicolumn{4}{c}{Fiducial cross-section, 2 jets } \\
\hline\hline
\pTb [\GeV{}] &  [25, 30]         & [30, 40]         & [40, 60]         & [60, 140]       \\
\hline
d$\sigma$/d\pTb  [nb/\GeV{}]                & 73  & 58  & 38  & 9.3 \\
\hline
 Statistical Uncertainty  (\%)                & 12   & 8   & 14  & 23  \\
 Systematic Uncertainty  (\%)                & 26 & 22   & 21  & 31  \\
\hline
Correlation coefficients of &  1      & 0.585  & $-0.45$  & $-0.08$  \\
statistical uncertainties  &  & 1      & 0.069  & $-0.29$  \\
&  &  & 1      & $-0.20$  \\
&  &  &  & 1      \\
\hline
Correlation coefficients of & 1      & 0.900  & 0.550  & 0.544 \\
systematic uncertainties  & & 1      & 0.795  & 0.719  \\
&  &  & 1      & 0.775  \\
&  &  &  & 1      \\
\hline\hline
\end{tabular}
\label{tabdiff2}
\end{center}
\end{table*}

The NLO predictions of \mcfm\ and \powheg\ are evaluated using the MSTW2008~\cite{MSTW2008andHessian} NLO PDF, and the following dynamic renormalization and factorization scales ($\mu_R$ and $\mu_F$) are chosen\footnote{
In the 5-flavour number scheme, the production of one $b$-jet in the final state with an associated 
light jet can occur. In those cases, one of the two last terms in  eqn.~\ref{eq:renorm} is omitted. 
}:
\begin{equation}
\mu_F^2=\mu_R^2=m^2_{\ell\nu}+\pt^2(\ell\nu)+\frac{m_b^2+\pt^2(b)}{2}+\frac{m_{\bar{b}}^2+\pt^2(\bar{b})}{2}. 
\label{eq:renorm}
\end{equation}
The PDF uncertainty is calculated with the MSTW2008 eigenvectors using the Hessian procedure~\cite{MSTW2008andHessian}. The dependence of the result on the choice of scale, which dominates the theoretical uncertainty, is evaluated by varying the scale conservatively between a quarter  and four times the value in equation~\ref{eq:renorm}, as in ref.~\cite{ref:4FNS5FNS}. These variations are used to calculate an asymmetric uncertainty before applying vetoes on additional jets. The effect of  jet vetoes is then taken into account following the procedure outlined in ref.~\cite{Tackmann}.

To compare the NLO calculations with data, the impact of non-perturbative effects and double-parton interactions has to be considered.
 The \mcfm\ predictions are only available at the parton level, while the \powheg\ predictions are interfaced with \pythia\   to model the non-perturbative effects of hadronization and the underlying event.
A multiplicative correction derived from the \powheg\ sample is therefore applied to the \mcfm\ calculation to account for these non-perturbative effects. The uncertainty on the hadronization component of this correction is estimated by comparing the \pythia\ and \herwig{} parton showers, while the uncertainty on the underlying event component is estimated using the alternative Perugia2011~\cite{isrfsr1} tune instead of the AUET2B~\cite{AUET2B} one. 
The effect of double-parton interactions, where a $W$ boson and heavy-flavour jet are produced from different parton--parton interactions within the same proton, also has to be considered. Neither the \mcfm\ nor the \powheg\ calculations include this contribution, therefore an additive correction derived from the \alpgen\ simulation interfaced to \herwig\ and \textsc{Jimmy} has beeen applied to both calculations. This correction represents a 25\% effect on the total cross-section, concentrated in the lowest momentum bins of the 1-jet region. The DPI contribution in \alpgen\ has been shown to agree at the detector level with the ATLAS measurement of $\sigma_{\rm eff}$ in the $W$+2-jet sample~\cite{DPICONF}. Based on this measurement, a 
$^{+39}_{-28}\%$ uncertainty is assigned to the DPI correction.
The non-perturbative and  DPI corrections for the 1-jet and 2-jets regions are presented in table~\ref{tab:nonPertCorr}. The fully corrected \mcfm\ predictions are presented in table~\ref{tab:theory} for the 1-jet, 2-jet and 1+2-jet fiducial regions.

\begin{table}
\caption{ Multiplicative correction factors for non-perturbative effects  and additive corrections for double-parton interactions, derived from the \alpgen\ simulation, applied to the  \mcfm\ and \powheg\ predictions for the comparisons with unfolded results. The non-perturbative uncertainties include the hadronization and underlying event modelling, while the DPI uncertainties are based on the ATLAS measurement of $\sigma_{\rm eff}$~\cite{DPICONF}.
} 
	\centering
\begin{tabular}{l|c|c}
\multicolumn{3}{c}{  }\\
\hline
Correction & 1 jet & 2 jets \\
\hline
Non-perturbative & $0.92\pm 0.02$ (had.) $\pm 0.03$ (UE)  & $0.96\pm 0.05$ (had.) $\pm 0.03$ (UE) \\
\hline
DPI  [pb] & $1.02\pm 0.05$ (stat) $^{+0.40}_{-0.29}$ (syst) & $0.32\pm 0.02$ (stat) $^{+0.12}_{-0.09}$ (syst) \\
\hline
\end{tabular}
\label{tab:nonPertCorr} 
\end{table}

\begin{table}
\caption{ Theoretical NLO predictions for the \Wbjet fiducial cross-section for one lepton flavour calculated with the \mcfm\ program, corrected for non-perturbative effects and DPI contributions. } 
	\centering
\begin{tabular}{l|c}
\multicolumn{2}{c}{  }\\
\hline
\multicolumn{2}{c}{  \mcfm\ NLO prediction [pb] }\\
\hline
1 jet & $3.01\pm 0.07$ (stat) $^{+0.72}_{-0.54}$ (scale) $\pm 0.04$ (PDF)   $\pm 0.08$ (non-pert)   $^{+0.40}_{-0.29}$ (DPI)  \\
\hline
2 jets & $1.69\pm 0.06$ (stat) $^{+0.40}_{-0.23}$ (scale) $\pm 0.04$ (PDF)  $\pm 0.08$ (non-pert)    $^{+0.12}_{-0.09}$ (DPI)  \\
\hline
1+2 jets &  $4.70\pm 0.09$ (stat) $^{+0.60}_{-0.49}$ (scale) $\pm 0.06$ (PDF)  $\pm 0.16$ (non-pert) $^{+0.52}_{-0.38}$ (DPI)  \\
\hline

\end{tabular}
\label{tab:theory} 
\end{table}

\clearpage
\section{Results without single-top subtraction}

The \Wbjet cross-section is also measured including the contribution of the single-top process.
These measurements provide a complementary perspective on the $W$+$b$-tagged-jet sample, and they have a higher statistical precision than the single-top subtracted ones, especially at high \pTb.

For each analysis region and \pTb bin, the same ML fit as for the \Wbjet measurement is used, as well as the same estimates and constraints for the multijet, \ttbar{}, $Z$+jets and diboson backgrounds. In the fit to the CombNN distribution, the \Wbjet and single-top templates are merged accounting for their respective predicted cross-sections, and they form a single template whose normalization is estimated.
As a consequence of the single-top process being considered as part of the signal, the number of nuisance parameters in the fit is reduced, thereby increasing its statistical precision.
After the CombNN fit, the number of estimated \Wbjet and single-top events is  unfolded to a common fiducial region, identical to the  \Wbjet fiducial region, using correction factors and a response matrix built from the sum of the two Monte Carlo samples.

The systematic uncertainties from the fit and unfolding steps are accounted for using the same methods as for the single-top subtracted measurement. An additional uncertainty is introduced to account for the relative normalization of \Wbjet and single-top. Alternative samples, in which the amounts of \Wbjet and single-top are doubled in turn, are used to perform the unfolding. The largest deviation obtained with respect to the nominal result, approximately 5\%, is then quoted as a separate systematic uncertainty.

The resulting fiducial cross-sections for \Wbjets plus single-top, combining the electron and muon channels, are:

$\sigma_\mathrm{fid}$ (1 jet) =  $5.9 \pm 0.2~\mathrm{(stat)}\pm 1.3~\mathrm{(syst)~pb}$, 
 
$\sigma_\mathrm{fid}$ (2 jet) = $3.7 \pm 0.1~\mathrm{(stat)}  \pm 0.8~\mathrm{(syst)~pb}$,
 
$\sigma_\mathrm{fid}$ (1+2 jet) = $9.6\pm 0.2~\mathrm{(stat)} \pm 1.7~\mathrm{(syst)~pb}$. 
 
 The corresponding expected cross-sections, calculated for the \Wbjets process using \alpgen\  interfaced to \herwig{} and \jimmy\ and scaled by the NNLO inclusive $W$ normalization factor  and for the single-top processes using \acermc\ interfaced to \pythia\ and scaled to NLO, are 3.6~pb, 3.0~pb and 6.6~pb, respectively. 
The differential results as a function of \pTb are presented in figure~\ref{Fig:resultsWbSt} and tables~\ref{tabdiff1st} and~\ref{tabdiff2st}.

\begin{figure}[!h]
    \begin{center}
  \includegraphics[width=0.49  \textwidth]{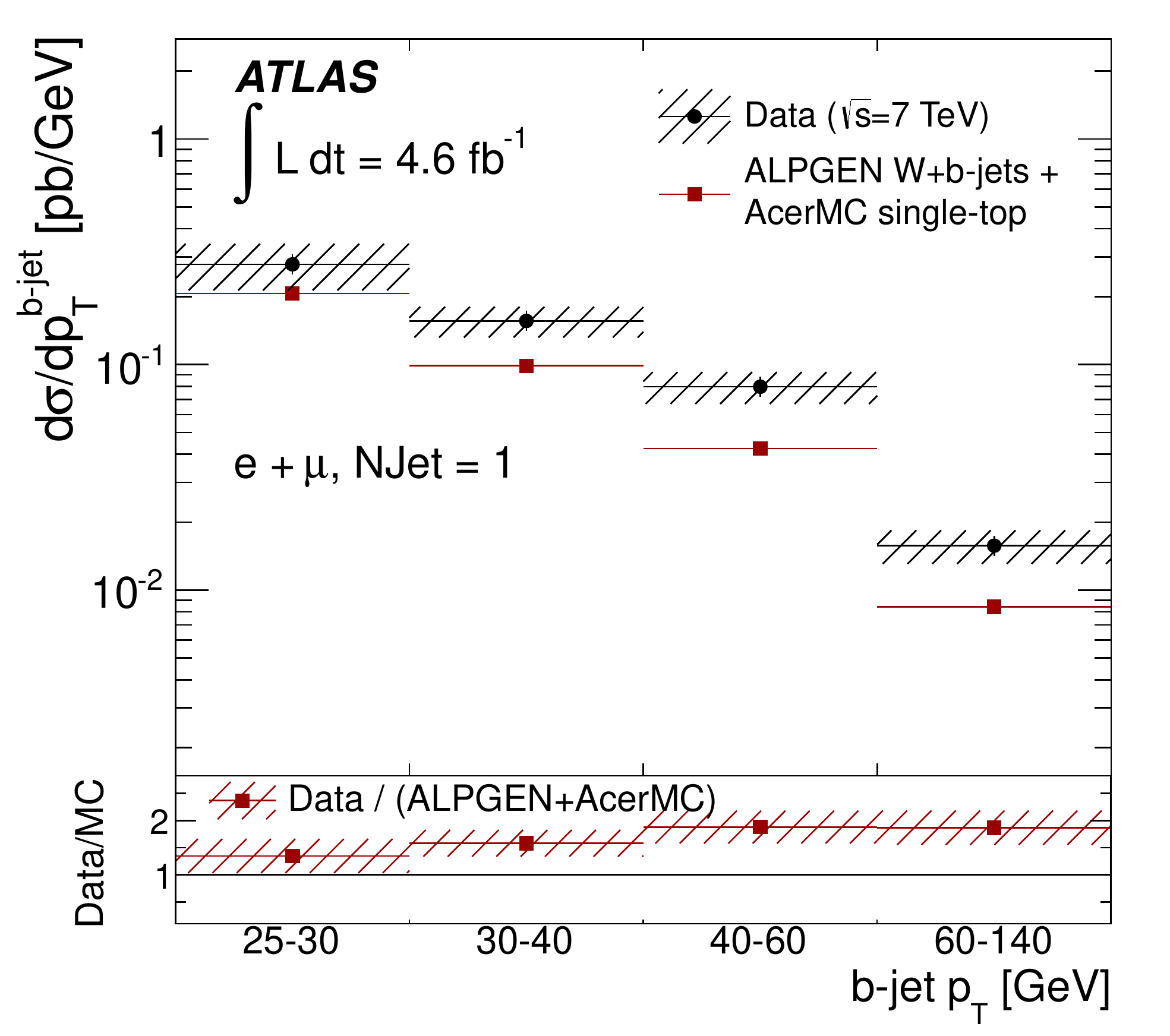}
    \includegraphics[width=0.49  \textwidth]{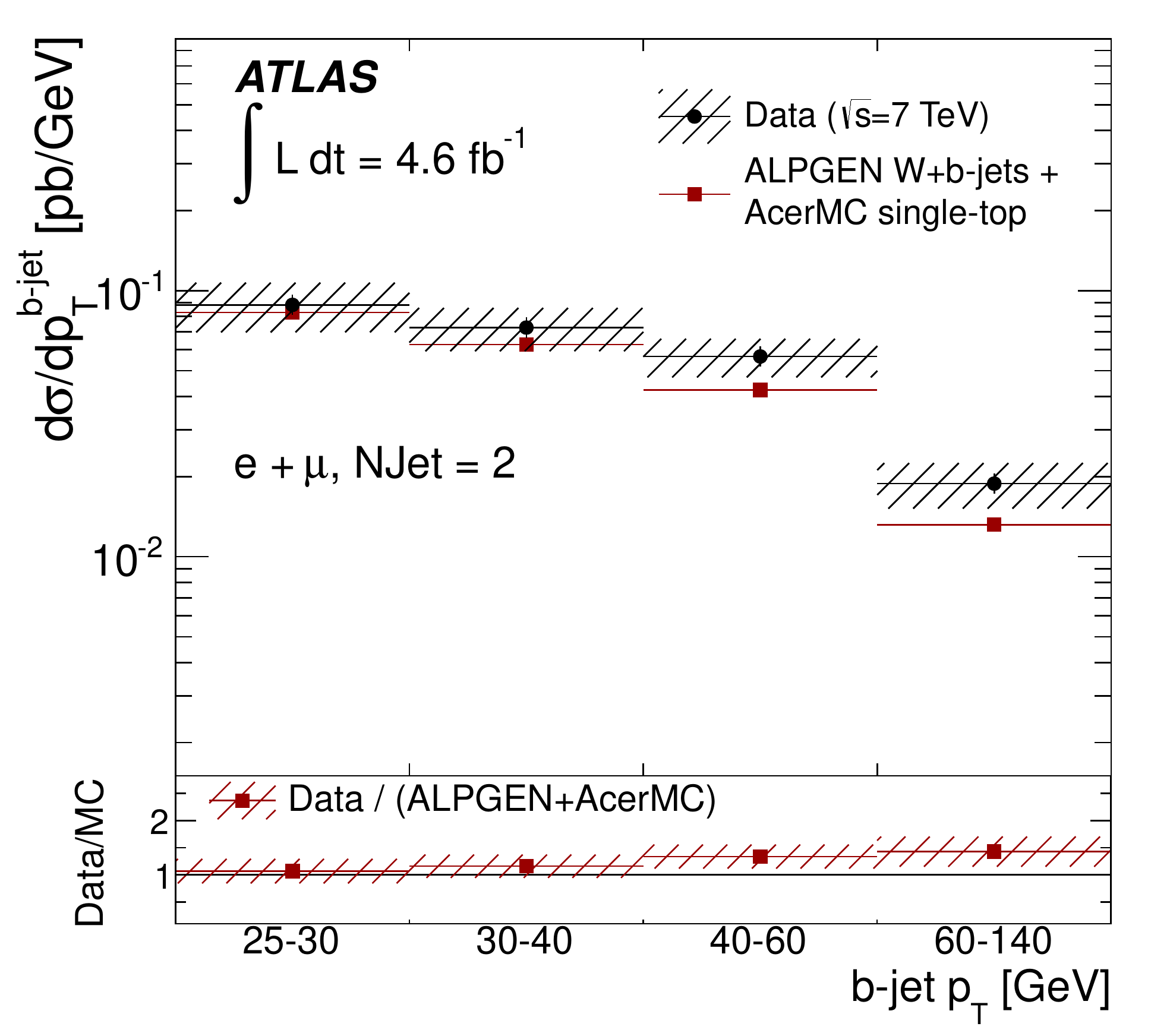}
        \caption{Measured differential \Wbjet cross-section without single-top subtraction as a function of \pTb in the 1-jet (left) and 2-jet (right) samples, obtained by combining the electron and muon channels. The measurements are compared to the \Wbjet plus single-top predictions obtained using \alpgen\  interfaced to \herwig{} and \jimmy\ and scaled by the NNLO inclusive $W$ normalization factor  plus \acermc\ interfaced to \pythia\ and scaled to the NLO single-top cross-section. The ratios between measured and predicted cross-sections are also shown.}
        \label{Fig:resultsWbSt}
    \end{center}
\end{figure}

\begin{table*}[ht]
\caption{Measured fiducial \Wbjets cross-section without single-top subtraction in the 1-jet  region, with statistical and systematic uncertainties
and their correlations in bins of \pTb.}
\begin{center}
\begin{tabular}{l c c c c}
\hline\hline
 \multicolumn{4}{c}{Fiducial cross-section  of \Wbjet $+$ single-top, 1 jet  } \\
\hline\hline
\pTb [\GeV{}] &  [25, 30]         & [30, 40]         & [40, 60]         & [60, 140]       \\
\hline
    d$\sigma$/d\pTb  [nb/\GeV{}]          & 278  & 156  & 80  & 15.7   \\
\hline
 Statistical Uncertainty  (\%)                & 6   & 4   & 5  & 5  \\
 Systematic Uncertainty  (\%)                & 23 & 15   & 15  & 16  \\
\hline
  Correlation coefficients of  & 1      & 0.401  & $-0.31$  & $-0.03$  \\
statistical uncertainties &  & 1      & 0.00  & $-0.13$  \\
&  &  & 1      & $-0.05$  \\
&  &  &  & 1      \\
\hline
 Correlation coefficients of  &   1      & 0.840  & 0.682  & 0.866  \\
systematic uncertainties &  & 1      & 0.935  & 0.875  \\
&  &  & 1      & 0.861  \\
&  &  &  & 1      \\
\hline\hline
\end{tabular}
\label{tabdiff1st}
\end{center}
\end{table*}

\begin{table*}[ht]
\caption{Measured fiducial \Wbjets cross-section without single-top subtraction in the 2-jet  region, with statistical and systematic uncertainties
and their correlations in bins of \pTb.}
\begin{center}
\begin{tabular}{l c c c c}
\hline\hline
 \multicolumn{4}{c}{Fiducial cross-section of \Wbjet $+$ single-top, 2 jets } \\
\hline\hline
\pTb [\GeV{}] &  [25, 30]         & [30, 40]         & [40, 60]         & [60, 140]       \\
\hline
    d$\sigma$/d\pTb  [nb/\GeV{}]             & 88  & 73  & 56.5  & 18.8  \\
\hline
 Statistical Uncertainty  (\%)                & 8   & 5   & 6  & 5  \\
 Systematic Uncertainty  (\%)                & 20 & 18   & 16  & 19  \\
\hline
 Correlation coefficients of & 1      & 0.602  & $-0.27$  & $-0.08$  \\
statistical uncertainties   &  & 1      & 0.125  & $-0.18$  \\
&  &  & 1      & $-0.12$  \\
&  &  &  & 1      \\
\hline
Correlation coefficients of  &  1      & 0.905  & 0.723  & 0.792  \\
systematic uncertainties &  & 1      & 0.925  & 0.940  \\
&  &  & 1      & 0.885  \\
&  &  &  & 1      \\
\hline\hline
\end{tabular}
\label{tabdiff2st}
\end{center}
\end{table*}

\clearpage
\section{Conclusions}

A measurement of the cross-section of $W$ boson production in association with $b$-jets at  $\sqrt{s} = 7 \TeV$ is presented, based on data corresponding to an integrated luminosity of 4.6~\ifb\ collected with the ATLAS detector at the LHC. 
The measurement is performed with a single $b$-tagged jet requirement in the $W$+1-jet and $W$+2-jets samples. \Wbjet yields are estimated separately in the electron and muon decay channel and unfolded to a common fiducial region. Good agreement is found between the results in the electron and muon channels and combined measurements are provided.

In the 1-jet  region, the measured fiducial cross-section is $5.0 \pm 0.5~\mathrm{(stat)}\pm 1.2~\mathrm{(syst)~pb}$, consistent within $1.5 \sigma$ with NLO predictions. In the 2-jet  region, the measured fiducial cross-section is  $2.2 \pm 0.2~\mathrm{(stat)} \pm 0.5~\mathrm{(syst)~pb}$, in good agreement with the theoretical calculations.
As a result, the combined 1+2-jet measurement, yielding a cross-section of  $7.1 \pm 0.5~\mathrm{(stat)}\pm 1.4~\mathrm{(syst)~pb}$, is found to be consistent within $1.5\sigma$ with the MCFM NLO prediction, corrected for hadronization and DPI effects, of $4.70 \pm 0.09~\mathrm{(stat)} ^{+0.60}_{-0.49}~\mathrm{(scale)} \pm 0.06 ~\mathrm{(PDF)} \pm 0.16$ (non-pert) $^{+0.52}_{-0.38}~\mathrm{(DPI)~pb}$.

A differential cross-section measurement as a function of the leading  \bjet \pT{} is also presented, for jets in the \pT\ range between 25~\GeV\ and 140~\GeV. In the 1-jet fiducial region, the measured cross-section is larger than the NLO predictions, but compatible within the theoretical and experimental uncertainties. The same measurement in the 2-jet fiducial region is found to be in agreement with the theoretical predictions.

A second set of measurements, including the single-top contribution, is also presented. 
In the 1-jet fiducial region, the measured \Wbjet plus single-top cross-section is $5.9 \pm 0.2~\mathrm{(stat)}\pm 1.3~\mathrm{(syst)~pb}$, while in the 2-jet fiducial region it is $3.7 \pm 0.1~\mathrm{(stat)}  \pm 0.8~\mathrm{(syst)~pb}$. The combined 1+2-jet fiducial cross-section is measured to be $9.6\pm 0.2~\mathrm{(stat)} \pm 1.7~\mathrm{(syst)~pb}$.
The corresponding \bjet \pT\ differential cross-sections have significantly reduced uncertainties with respect to the single-top subtracted measurements and can be compared to combined single top-quark and \Wbjet calculations in the future. 

\acknowledgments
We thank CERN for the very successful operation of the LHC, as well as the
support staff from our institutions without whom ATLAS could not be
operated efficiently.

We acknowledge the support of ANPCyT, Argentina; YerPhI, Armenia; ARC,
Australia; BMWF and FWF, Austria; ANAS, Azerbaijan; SSTC, Belarus; CNPq and FAPESP,
Brazil; NSERC, NRC and CFI, Canada; CERN; CONICYT, Chile; CAS, MOST and NSFC,
China; COLCIENCIAS, Colombia; MSMT CR, MPO CR and VSC CR, Czech Republic;
DNRF, DNSRC and Lundbeck Foundation, Denmark; EPLANET, ERC and NSRF, European Union;
IN2P3-CNRS, CEA-DSM/IRFU, France; GNSF, Georgia; BMBF, DFG, HGF, MPG and AvH
Foundation, Germany; GSRT and NSRF, Greece; ISF, MINERVA, GIF, DIP and Benoziyo Center,
Israel; INFN, Italy; MEXT and JSPS, Japan; CNRST, Morocco; FOM and NWO,
Netherlands; BRF and RCN, Norway; MNiSW, Poland; GRICES and FCT, Portugal; MERYS
(MECTS), Romania; MES of Russia and ROSATOM, Russian Federation; JINR; MSTD,
Serbia; MSSR, Slovakia; ARRS and MVZT, Slovenia; DST/NRF, South Africa;
MICINN, Spain; SRC and Wallenberg Foundation, Sweden; SER, SNSF and Cantons of
Bern and Geneva, Switzerland; NSC, Taiwan; TAEK, Turkey; STFC, the Royal
Society and Leverhulme Trust, United Kingdom; DOE and NSF, United States of
America.

The crucial computing support from all WLCG partners is acknowledged
gratefully, in particular from CERN and the ATLAS Tier-1 facilities at
TRIUMF (Canada), NDGF (Denmark, Norway, Sweden), CC-IN2P3 (France),
KIT/GridKA (Germany), INFN-CNAF (Italy), NL-T1 (Netherlands), PIC (Spain),
ASGC (Taiwan), RAL (UK) and BNL (USA) and in the Tier-2 facilities
worldwide.


\onecolumn
\clearpage 

\begin{flushleft}
{\Large The ATLAS Collaboration}

\bigskip

G.~Aad$^{\rm 48}$,
T.~Abajyan$^{\rm 21}$,
B.~Abbott$^{\rm 111}$,
J.~Abdallah$^{\rm 12}$,
S.~Abdel~Khalek$^{\rm 115}$,
A.A.~Abdelalim$^{\rm 49}$,
O.~Abdinov$^{\rm 11}$,
R.~Aben$^{\rm 105}$,
B.~Abi$^{\rm 112}$,
M.~Abolins$^{\rm 88}$
% ...
\bigskip
\\
% ...
$^{11}$ Institute of Physics, Azerbaijan Academy of Sciences, Baku, Azerbaijan\\
$^{12}$ Institut de F{\'\i}sica d'Altes Energies and Departament de F{\'\i}sica de la Universitat Aut{\`o}noma de Barcelona and ICREA, Barcelona, Spain\\
% ...
$^{21}$ Physikalisches Institut, University of Bonn, Bonn, Germany\\
% ...
$^{38}$ AGH University of Science and Technology, Faculty of Physics and Applied Computer Science, Krakow, Poland\\
% ...
$^{48}$ Fakult{\"a}t f{\"u}r Mathematik und Physik, Albert-Ludwigs-Universit{\"a}t, Freiburg, Germany\\
$^{49}$ Section de Physique, Universit{\'e} de Gen{\`e}ve, Geneva, Switzerland\\
% ...
$^{52}$ II Physikalisches Institut, Justus-Liebig-Universit{\"a}t Giessen, Giessen, Germany\\
% ...
$^{88}$ Department of Physics and Astronomy, Michigan State University, East Lansing MI, United States of America\\
% ...
$^{105}$ Nikhef National Institute for Subatomic Physics and University of Amsterdam, Amsterdam, Netherlands\\
% ...
$^{111}$ Homer L. Dodge Department of Physics and Astronomy, University of Oklahoma, Norman OK, United States of America\\
$^{112}$ Department of Physics, Oklahoma State University, Stillwater OK, United States of America\\
% ...
$^{115}$ LAL, Universit{\'e} Paris-Sud and CNRS/IN2P3, Orsay, France\\
% ...
$^{s}$ Also at Manhattan College, New York NY, United States of America\\
% ...
\end{flushleft}



\begin{thebibliography}{99}


\bibitem{MCFM}
J.~M. Campbell and R.~Ellis, {\em MCFM for the Tevatron and the LHC\/},
  {\em Nucl.\ Phys.\ Proc.\ Suppl.} {\bf 205-206} (2010)  10.

\bibitem{Powheg}
S.~Alioli, P.~Nason, C.~Oleari, and E.~Re, {\em {A general framework for
  implementing NLO calculations in shower Monte Carlo programs: the POWHEG
  BOX}\/},  
    {\em JHEP} {\bf 1006} (2010)  043 [arXiv:1002.2581].

\bibitem{MC_at_NLO}
 S. Frixione and B. R. Webber, {\em Matching NLO QCD computations and parton shower
  simulations}, {\em JHEP}  {\bf 06} (2002) 029 [hep-ph/0204244]; \\ 
  S. Frixione, P. Nason and B. R. Webber, {\em Matching NLO QCD and parton showers in heavy flavour production}, {\em JHEP}  {\bf 08 } (2003) 007 [hep-ph/0305252]; \\
  S. Frixione, E. Laenen and P. Motylinski, {\em Single-top production
  in MC@NLO}, {\em JHEP}  {\bf 03 } (2006) 092 [hep-ph/0512250].

\bibitem{ref:4FNS5FNSref}
 S. Badger, J. M. Campbell and R. K. Ellis, {\em QCD corrections to the hadronic production of a heavy quark pair and a W-boson including decay correlations}, {\em JHEP} {\bf 1103} (2011) 027  [arXiv:1011.6647]; \\
  F. F. Cordero, L. Reina and D. Wackeroth, {\em Associated production of a W or Z boson with bottom quarks at the Tevatron and the LHC},  {\em PoS} {\bf
  RADCOR2009} (2010) 055  [arXiv:1001.3362]; \\
  J.~M.~Campbell, R.~K.~Ellis, F.~Febres Cordero, F.~Maltoni, L.~Reina, D.~Wackeroth and S.~Willenbrock, {\em Associated Production of a W Boson and One b Jet, Phys. Rev.} {\bf D 79} (2009) 034023  [arXiv:0809.3003].
    
  \bibitem{ref:4FNS5FNS} 
  J.~M.~Campbell, F.~Caola, F.~Febres Cordero, L.~Reina and D.~Wackeroth,
{\em NLO QCD predictions for $W+1$ jet and $W+2$ jet production with at least one $b$ jet at the 7 TeV LHC, 
  Phys.\ Rev.} {\bf D 86}  (2012) 034021 [arXiv:1107.3714].
  
\bibitem{MPI_WBB} 
  E.~L.~Berger, C.~B.~Jackson, S.~Quackenbush and G.~Shaughnessy,
{\em Calculation of W\bbbar\ Production via Double Parton Scattering at the LHC,
  Phys.\ Rev.} {\bf D 84} (2011) 074021  [arXiv:1107.3150].
  
  \bibitem{WH}
ATLAS Collaboration, \emph{Search for the Standard Model Higgs boson produced in association with a vector boson and decaying to a b-quark pair with the ATLAS detector}, {\em Phys. Lett. }{\bf B 718} (2012) 369 [arXiv:1207.0210].


\bibitem{HiggsComb}
ATLAS Collaboration, {\em Observation of a new particle in the search for the Standard Model Higgs boson with the ATLAS detector at the LHC, Phys. Lett.} {\bf B 716} (2012) 1 [arXiv:1207.7214].

\bibitem{NewPhys1} 
 ATLAS Collaboration,
 \emph{Search for light top squark pair production in final states with leptons and b-jets with the ATLAS detector in $\sqrt{s} = 7$ \TeV\  proton-proton collisions}, submitted to {\em Phys. Lett.} {\bf B} (2012) [arXiv:1209.2102].

\bibitem{SingleTop}
  ATLAS Collaboration,
  \emph{Measurement of the t-channel single top-quark production cross section in pp collisions at $\sqrt{s} = 7$ \TeV\  with the ATLAS detector},
{\em Phys. Lett.} {\bf B 717} (2012) 330 [arXiv:1205.3130].

  
  \bibitem{Aaltonen:2009qi}
CDF Collaboration, T. Aaltonen et al, {\em First Measurement of the b-jet Cross Section in Events with a W Boson in $p\bar{p}$ Collisions at $\sqrt{s}$ = 1.96 TeV},
{\em Phys.\ Rev.\ Lett.} {\bf 104} (2010) 131801  [arXiv:0909.1505].

\bibitem{D0Wb} 
  D0 Collaboration, V. Abazov et al,
{\em Measurement of the $p\bar{p}$ $\rightarrow$ W+b+X production cross section at $\sqrt{s}$ =1.96 TeV} [arXiv:1210.0627].

\bibitem{bib:ATLASWb}
{ATLAS} Collaboration, {\em Measurement of the cross section for the
  production of a W boson in association with b-jets in pp collisions at
  $\sqrt{s} = 7$ \TeV\  with the ATLAS detector}, {\em Phys. Lett. }{\bf B 707} (2012)
  418 
  [arXiv:1109.1470].

\bibitem{bib:JINST}
 ATLAS Collaboration, {\em The ATLAS Experiment at the CERN Large Hadron Collider},  {\em JINST} {\bf 3} (2008) S08003.

\bibitem{ALPGEN}
 M. L. Mangano, M. Moretti, F. Piccinini, R. Pittau and A. D. Polosa, {\em ALPGEN, a
  generator for hard multiparton processes in hadronic collisions}, {\em JHEP} {\bf 07}
  (2003) 001.

\bibitem{Herwig}
 G. Corcella et al., {\em HERWIG 6.5: an event generator for Hadron Emission
  Reactions With Interfering Gluons (including supersymmetric processes)}, {\em JHEP}
  {\bf 01} (2001) 010; \\ G. Corcella et al., {\em HERWIG 6.5 release notes} [hep-ph/0210213].
  
  \bibitem{Jimmy}
  J. M. Butterworth, J. R. Forshaw, and M. H. Seymour, {\em Multiparton Interactions in Photoproduction at HERA}, {\em Z. Phys.}  {\bf C 72} (1996) 637 [hep-ph/9601371].
   
\bibitem{MLM}
M.~L.~Mangano and R.~Pittau, {\em {Multijet matrix elements and shower
  evolution in hadronic collisions: Wbb + n jets as a case study}\/},  {\em Nucl.
  Phys. } {\bf B 632} (2002)  343 
  [hep-ph/0108069].
  
  \bibitem{Sjostrand:2006za}
T.~Sjostrand, S.~Mrenna, and P.~Z. Skands, {\em {PYTHIA 6.4 Physics and
  Manual}\/},  
    {\em JHEP} {\bf  0605} (2006)  026
[hep-ph/0603175].

\bibitem{bib:ACERMC}
 B.P. Kersevan and E. Richter-Was, {\em The Monte Carlo event generator AcerMC
  version 2.0 with interfaces to PYTHIA 6.2 and HERWIG 6.5}
  [hep-ph/0405247].

\bibitem{FEWZ}
 C. Anastasiou, L. Dixon, K. Melnikov and F. Petriello, 
 {\em  High-precision QCD at hadron colliders: electroweak gauge boson rapidity distributions at NNLO},
 {\em Phys. Rev.} {\bf D 69} (2004)
  094008  [hep-ph/0312266].

\bibitem{VVnormalization}
  J.~M.~Campbell, R.~K.~Ellis and C.~Williams,
  {\em Vector boson pair production at the LHC},
  {\em JHEP} {\bf 1107} (2011) 018
  [arXiv:1105.0020].

\bibitem{SingleTopNormalization}
  J.~M.~Campbell, R.~K.~Ellis and F.~Tramontano,
  {\em Single top production and decay at next-to-leading order},
 {\em Phys. Rev.} {\bf D  70} (2004) 094012
  [hep-ph/0408158].

\bibitem{TopNormalization}
  M.~Aliev, H.~Lacker, U.~Langenfeld, S.~Moch, P.~Uwer and M.~Wiedermann,
  {\em HATHOR: HAdronic Top and Heavy quarks crOss section calculatoR},
  {\em Comput.\ Phys.\ Commun.}  {\bf 182} (2011) 1034
  [arXiv:1007.1327].

  \bibitem{AtlasSimulation}
  ATLAS Collaboration, {\em The ATLAS Simulation Infrastructure}, {\em Eur. Phys. J.}  {\bf C 70} (2010) 823 [arXiv:1005.4568].

\bibitem{Geant4} S. Agostinelli et. al., \emph{GEANT4: A simulation toolkit}, \emph{Nucl. Instrum. Meth.} {\bf A 506} (2003) 250.

\bibitem{bib:Lum1}
{ATLAS} Collaboration,  {\em {Luminosity determination in pp
  collisions at $\sqrt{s} = 7$ \TeV\  using the ATLAS detector at the LHC}\/},
  {\em Eur. Phys. J.} {\bf C 71} (2011)  1630
[arXiv:1101.2185].

\bibitem{bib:Lum2}
{ATLAS Collaboration}, {\em Luminosity Determination in pp Collisions at
  $\sqrt{s} = 7$ \TeV\  using the ATLAS Detector in 2011\/},  
  \href{http://cdsweb.cern.ch/record/1376384}{ ATLAS-CONF-2011-116}.

\bibitem{TagProbeEle}
{ATLAS Collaboration}, {\em Electron performance measurements with the ATLAS detector using the 2010 LHC proton-proton collision data}, 
 {\em Eur. Phys. J.} {\bf C 72} (2012) 1909 [arXiv:1110.3174].

\bibitem{MomentumScaleMu}
{ATLAS Collaboration}, {\em Muon reconstruction efficiency in reprocessed 2010 LHC proton-proton collision data recorded with the ATLAS detector},
\href{http://cdsweb.cern.ch/record/1345743}{ ATLAS-CONF-2011-063}.

\bibitem{TagProbeMu}
{ATLAS Collaboration}, {\em ATLAS muon momentum resolution in the first pass reconstruction of the 2010 pp Collision Data at $\sqrt{s} = 7 \TeV$}, 
\href{http://cdsweb.cern.ch/record/1338575}{ ATLAS-CONF-2011-046}.

\bibitem{TopoClusters}
W.~Lampl et al., {\em Calorimeter clustering algorithms: description and
  performance},  
  \href{http://cdsweb.cern.ch/record/1099735}{ ATL-LARG-PUB-2008-002}.

\bibitem{ref:antikt}
M.~Cacciari, G.~P. Salam, and G.~Soyez, {\em The anti-k$_t$ jet clustering
  algorithm\/},  {\em JHEP} {\bf 04} (2008)  063
[arXiv:0802.1189].

\bibitem{JetCleaning}
ATLAS Collaboration, {\em Selection of jets produced in proton-proton collisions with the ATLAS detector using 2011 data},
\href{http://cdsweb.cern.ch/record/1430034}{ ATLAS-CONF-2012-020}.

\bibitem{ref:JES}
 ATLAS Collaboration, {\em Jet energy measurement with the ATLAS detector in proton-proton collisions at $\sqrt{s} = 7$ \TeV\ }, submitted to {\em Eur. Phys. J.}  
 [arXiv:1112.6426].
 
 \bibitem{jeseta}
 ATLAS Collaboration, 
 {\em In situ jet pseudorapidity intercalibration of the ATLAS detector using dijet events in $\sqrt{s} = 7 \TeV$ proton-proton 2011 data}, 
 \href{http://cdsweb.cern.ch/record/1474490}{ ATLAS-CONF-2012-124}.
 
  \bibitem{jes}
 ATLAS Collaboration, 
 {\em Light-quark and Gluon Jets: Calorimeter Response, Jet Energy Scale Systematics and Properties}, 
\href{http://cdsweb.cern.ch/record/1480629}{ ATLAS-CONF-2012-138}.
 
 \bibitem{JESpileup}
 ATLAS Collaboration,  
 {\em Pile-up corrections for jets from proton-proton collisions at $\sqrt{s} = 7 \TeV$ in ATLAS in 2011},
 \href{http://cdsweb.cern.ch/record/1459529} { ATLAS-CONF-2012-064}.

\bibitem{JESInSituZ}
 ATLAS Collaboration,  
  {\em Probing the measurement of jet energies with the ATLAS detector using $Z$+jet events from proton-proton collisions at $\sqrt{s} = 7 \TeV$},
  \href{http://cdsweb.cern.ch/record/1452641}{ ATLAS-CONF-2012-053}.
 
 \bibitem{JESInSituGamma}
 ATLAS Collaboration,    
  {\em Probing the measurement of jet energies with the ATLAS detector using photon+jet events in proton-proton collisions at $\sqrt{s} = 7 \TeV$},
 \href{http://cdsweb.cern.ch/record/1459528} { ATLAS-CONF-2012-063}.

\bibitem{bib:AdvancedTaggers}
The ATLAS collaboration, 
{\em Commissioning of the ATLAS high-performance
  $b$-tagging algorithms in the $7 \TeV$ collision data},
   \href{http://cdsweb.cern.ch/record/1369219}{ ATLAS-CONF-2011-102}.

\bibitem{ref:BtaggingSF}
{ATLAS} Collaboration, 
{\em Measurement of the b-tag Efficiency in a Sample of
  Jets Containing Muons with 5 \ifb\ of Data from the ATLAS Detector},
 \href{http://cdsweb.cern.ch/record/1435197}{ ATLAS-CONF-2012-043}.

\bibitem{ref:BtaggingSFC}
{ATLAS} Collaboration, 
{\em \bjet tagging calibration on \cjets containing $D^{*+}$ mesons},
 \href{http://cdsweb.cern.ch/record/1435193}{ ATLAS-CONF-2012-039}.
 
 \bibitem{ref:BtaggingSFL}
{ATLAS} Collaboration, 
{\em Measurement of the Mistag Rate with 5 \ifb\ of Data Collected by the ATLAS Detector},
 \href{http://cdsweb.cern.ch/record/1435194}{ ATLAS-CONF-2012-040}.

\bibitem{ref:METRefFinal}
 ATLAS Collaboration,  
  {\em Reconstruction and Calibration of Missing Transverse Energy and Performance in Z and W events in ATLAS Proton-Proton Collisions at 7 \TeV},
 \href{http://cdsweb.cern.ch/record/1355703}{ ATLAS-CONF-2011-080}.

\bibitem{Barillari:2009zza}
T.~Barillari et al., {\em Local hadronic calibration},  
\href{http://cdsweb.cern.ch/record/1112035} { ATL-LARG-PUB-2009-001}.

\bibitem{ztheo}
K.~Melnikov, F.~Petriello,
{\em Electroweak gauge boson production at hadron colliders through O$(\alpha_{s}^{2})$},
{\em Phys. Rev.} {\bf D 74} (2006) 114017
[hep-ph/0609070].

\bibitem{zjetatlas}
ATLAS Collaboration, 
{\em Measurement of the production cross section for $Z/\gamma^*$ in association with jets in pp collisions at $\sqrt{s} = 7 \TeV\ $ with the ATLAS detector},
{\em Phys. Rev.}   {\bf D 85} (2012) 032009 
[arXiv:1111.2690].

\bibitem{wwz1}
K.~Hagiwara, S.~Ishihara, R.~Szalapski, and D.~Zeppenfeld,
{\em Low energy effects of new interactions in the electroweak boson sector},
{\em Phys. Rev.}   {\bf D 48} (1993) 2182.

\bibitem{wwz2}
J. M.~Campbell and R. K.~Ellis,
{\em An update on vector boson pair production at hadron colliders},
{\em Phys. Rev.}   {\bf D 60} (1999) 113006.

\bibitem{PDG}
J. Beringer et al. (Particle Data Group), {\em The Review of Particle Physics}, {\em Phys. Rev.}   {\bf D 86} (2012) 010001.

\bibitem{iterDAgostini}
G.~D'Agostini, {\em {Improved Iterative Bayesian Unfolding}\/}
  [arXiv:1010.0632].

\bibitem{isrfsr1}
P. Skands, 
{\em Tuning Monte Carlo generators: The Perugia tunes},
{\em Phys. Rev.}   {\bf D 82} (2010) 074018 
[arXiv:1005.3457].

\bibitem{CTEQ6L1}
  J.~Pumplin, D.~R.~Stump, J.~Huston, H.~L.~Lai, P.~M.~Nadolsky and W.~K.~Tung,
{\em New generation of parton distributions with uncertainties from global QCD analysis},
  {\em JHEP} {\bf 0207} (2002) 012
  [hep-ph/0201195].

\bibitem{MRST2002}
  A.~D.~Martin, R.~G.~Roberts, W.~J.~Stirling and R.~S.~Thorne,
  {\em Uncertainties of predictions from parton distributions. 1: Experimental errors},
  {\em Eur.\ Phys.\ J.}  {\bf C 28} (2003) 455
  [hep-ph/0211080].

\bibitem{METuncert}
ATLAS Collaboration, {\em Performance of Missing Transverse Momentum Reconstruction in ATLAS with 2011 Proton-Proton Collisions at $\sqrt{s} = 7 \TeV$},
\href{http://cdsweb.cern.ch/record/1463915}{ ATLAS-CONF-2012-101}.

\bibitem{MSTW2008andHessian}
A.~Sherstnev and R.~S. Thorne, {\em {Parton distributions for LO
  generators}\/},  {\em Eur. Phys. J.} {\bf C 55} (2008)  553
[arXiv:0711.2473].

\bibitem{Tackmann}
I.~W. Stewart and F.~J. Tackmann, {\em {Theory Uncertainties for Higgs and
  Other Searches Using Jet Bins}\/},
   {\em Phys. Rev.} {\bf D 85} (2012)  034011
[arXiv:1107.2117].

\bibitem{AUET2B}
 ATLAS Collaboration,  
  {\em ATLAS tunes of PYTHIA 6 and Pythia 8 for MC11},
 \href{http://cdsweb.cern.ch/record/1363300}{ ATL-PHYS-PUB-2011-009}.

\bibitem{DPICONF}
ATLAS Collaboration, {\em Measurement of hard double-parton interactions
  in $W\rightarrow \ell\nu$ + 2 jet events at $\sqrt{s} = 7 \TeV$ with the ATLAS detector},
submitted to {\em NPJ} (2013) [arXiv:1301.6872].


\end{thebibliography}
\end{document}